\documentclass[a4paper]{article}

\usepackage{graphicx}
\usepackage{authblk} 				
\usepackage{amssymb,amsmath}		
	\DeclareMathOperator\arccot{arccot}
	\renewcommand\Re{\operatorname{Re}}
	\renewcommand\Im{\operatorname{Im}}
\usepackage[inline]{enumitem}  		
\usepackage{hyperref}				
\usepackage{multirow}				
\usepackage{natbib}          			
\usepackage{setspace} 				
\begin{document}
%
%
\title{\Large Appraisal of the Magnetotelluric Galvanic Electric Distortion by Optimisation of the Relation between Amplitude and Phase Tensors}
%
%
	\author[$\dagger$]{Maik Neukirch} 
	\affil[$\dagger$]{presently: independent researcher; previously: Barcelona Center for Subsurface Imaging, Institut de Ci\`encies del Mar, Barcelona, Spain}

	\author[$*$]{Xavier Garcia}
	\affil[$*$]{Barcelona Center for Subsurface Imaging, Institut de Ci\`encies del Mar, Barcelona, Spain}
	
	\author[$\dagger$]{Savitri Galiana} 

%
%
{
\maketitle
\vspace{-.75cm}
{\begin{center} 
Short title: Appraisal of Galvanic Electric Distortion
\end{center}}
\begin{abstract}
The introduction of the Phase Tensor marked a major breakthrough in the understanding, analysis and treatment of galvanic distortion of the electric field in the Magnetotelluric (MT) method. 
We build upon a recently formulated impedance tensor decomposition into the known Phase Tensor and an Amplitude Tensor that is shown to be complementary and algebraically independent of the Phase Tensor. 
This recent decomposition demonstrates that the Amplitude Tensor contains inductive and galvanic information of the subsurface and that the inductive information is physically coupled to the one contained in the Phase Tensor.
It is also demonstrated that, through this coupling, galvanic effects can be separated from inductive effects in the Amplitude Tensor. 
In this work we present an algorithm that employs this last finding to show that the MT galvanic electric distortion tensor can be separated from the inductive Amplitude Tensor and hence, that this distortion can be recovered for any given data up to a single constant usually denoted as galvanic shift or site gain. 
Firstly, to illustrate distortion effects on the Amplitude Tensor, we manually apply distortion by matrix multiplication to synthetic impedance tensor data. 
Then, we use the observations of that analysis to define an objective function, which minimises when there is no distortion present in the Amplitude Tensor. 
Secondly, we describe our algorithm that employs a genetic algorithm to find the optimal distortion tensor needed to correct the Amplitude Tensor, and therewith the impedance tensor. 
Lastly, we test the performance of the proposed methodology on synthetic data of known distortion, on a large scale (144 sites) synthetic data set of random distortion and on four real data sets taken from the BC87 data set that are reported to contain 3D inductive effects. 
The real data sets, \emph{lit007}/\emph{lit008} and \emph{lit901}/\emph{lit902}, demonstrate the utility of the proposed algorithm by revealing geological expected results in the impedance data for the first time. This could not be achieved before by alternative methods due to their inherent assumption of a 2D regional impedance, which is not required in our scheme. 
{\small {\bf 
\newline keywords: Galvanic Electric Distortion, MT Phase Tensor, MT Amplitude Tensor, Distortion Analysis, Tensor Decomposition} }
\end{abstract}
\tableofcontents
}
%
%


\section{Introduction}
In recent years, magnetotelluric (MT) 3D data inversion has become a feasible technique due to the growing choice of publicly available inversion codes \citep{siripunvaraporn:2005,siripunvaraporn:2009,kelbert:2014,avdeeva:2015} to the academic community. However, galvanic distortion of the electric field remains a recognised complication for MT 3D data and, if present, can lead to inversion artefacts and incorrect interpretation of obtained subsurface images \citep[][and references therein]{jones:2012b}. Analytic correction of MT data for galvanic electric distortion has been addressed for 1D regional impedances \citep{Larsen:1977} and for 2D regional impedances \citep{Bahr:1988,Groom:1989,Chave:1994,McNeice:2001,Bibby:2005}. Though successful appraisal of electric distortion has been reported by incorporating the distortion parameters into the model space of a 3D inversion \citep{avdeeva:2015}, to date no method is known to appreciate electric distortion for 3D data when 3D inversions with distortion parameters are not feasible due to a low number of stations or generally insufficient spatial coverage, i.e.~by data on a profile line.

Galvanic distortion is caused by near-surface inhomogeneities below the size of the experiment's resolution defined by its sampling rate \citep[][and references therein]{Larsen:1977,Bahr:1988,Groom:1989,Jiracek:1990}. Distortion effects can be classified as electric and magnetic field distortions, of which electric field distortion is constant over frequency and the magnetic field is decreasingly distorted with decreasing frequencies \citep{Chave:1994}. Magnetic field distortion decreases with increasing period and often only affects very short periods. Therefore, it is commonly neglected and it is also not the subject of this paper. Electric field distortion is described by a real-valued, frequency independent tensor $\mathbf{C}$ which relates the observed MT impedance tensor, $\mathbf{Z}_d$, to the regional MT impedance tensor, $\mathbf{Z}$, in absence of distortion: 
\begin{equation} 
\mathbf{Z}_d(\omega)=\mathbf{C} \mathbf{Z}(\omega), 
\end{equation}
where $\omega$ denotes frequency dependence \citep{Bahr:1988,Groom:1989}. In order to avoid misinterpretation due to distortion effects in measured data, \cite{Caldwell:2004} proposed the (electric) distortion-free MT Phase Tensor (PT): 
 \begin{equation} 
\mathbf{\Phi}=(\Re \mathbf{Z}_d)^{-1} \Im \mathbf{Z}_d=(\Re \mathbf{Z})^{-1} \mathbf{C}^{-1} \mathbf{C}\Im \mathbf{Z}=(\Re \mathbf{Z})^{-1}\Im \mathbf{Z},
\end{equation} 
as a relation between the real and imaginary parts of the impedance. The PT can be described by the determinant and three geometric parameters: strike angle, skew angle and anisotropy, all of which can be used to interpret MT data and to infer the underlying subsurface dimensionality free of (electric) distortion \citep{Booker:2014}. The PT is dimensionless and only indicates changes of subsurface conductivity \citep{Caldwell:2004} but recovery of the absolute values, by inversion, may be possible when several sites are available with sufficiently overlapping information and if the a-priori model is chosen appropriately \citep{patro:2012,Tietze:2015}, however, the authors acknowledge that (undistorted) amplitude information would be a great asset. 

\cite{Neukirch:2016a} propose to extend the PT formalism to an impedance tensor decomposition by defining the positive-definite, real-valued MT Amplitude Tensor (AT), $\mathbf{P}$ (read: capital greek letter $\rho$): 
 \begin{equation} 
\mathbf{Z}=\mathbf{P} e(\mathbf{\Phi}),
\label{eq:AmplitudePhaseDecomposition}
\end{equation} 
where the PT function is 
$e(\mathbf{\Phi})=\left(\sqrt{\left(\mathbb{I}+\mathbf{\Phi\Phi}^T\right)}\right)^{-1} + i \left(\sqrt{\left(\mathbb{I}+\mathbf{\Phi\Phi}^T\right)}\right)^{-1} \mathbf{\Phi}$, $\mathbb{I}$ is the unity matrix, $T$ denotes the matrix transpose and the square root symbol represents the matrix square root. The AT contains all impedance amplitude information and all electric distortion. It can be parameterised in three dimensionless geometric parameters: strike angle, skew angle and anisotropy, and one positive scalar (determinant), just like the PT. Furthermore, \cite{Neukirch:2016a} postulated the hypothesis that any unique subsurface geometry must link the geometric parameters of the inductive AT part and the ones of the exclusively inductive PT, and that this relationship can be leveraged to estimate inductive and galvanic parts of the AT.

In this work, we present a new methodology to estimate the undistorted regional MT impedance and the galvanic electric distortion tensor up to the site gain, based on the Amplitude-Phase decomposition and imposing the similarity between the geometric parameters of the AT and the PT. 
First, we exemplify how galvanic electric distortion affects the AT parameters and show that only for undistorted AT responses, PT and AT parameters are similar. 
We use the observation of the required similarity between (inductive) AT and PT geometric parameters to define an objective function that embodies this relationship and reaches a minimum, when a-priori guessed distortion parameters result in most similar AT and PT parameters and thus, best represent galvanic electric distortion effects. Then, an optimisation algorithm (here a genetic algorithm) systematically seeks the optimal distortion parameters to successfully remove electric galvanic distortion from the AT and therewith, from the impedance tensor. 
Lastly, we illustrate the performance of the proposed methodology on synthetic data of known distortion, on a large scale (144 sites) synthetic data set of random distortion \citep{Miensopust:2013} and on four real data sets taken from the BC87 data set that are reported to contain 3D inductive effects \citep{Jones:1988,Jones:1993a,Ledo:2001}. 
To verify our algorithm, we compare our results to the ones of the widely used distortion and strike angle analysis program \emph{strike} by \cite{McNeice:2001}, which assumes two-dimensionality for the regional impedances.
The real data sets, \emph{lit007}/\emph{lit008} and \emph{lit901}/\emph{lit902}, demonstrate the utility of the proposed methodology by revealing geological expected results in the impedance data for the first time. This could not be achieved before by alternative methods due to their inherent assumption of a 2D regional impedance, which is not required in our scheme. 

The manuscript is structured as follows: Firstly, we outline the whole algorithm, secondly, we describe each step in detail and argue for the specific design choices, then it follows the verification on various examples, including specifically 3D data to demonstrate that the approach makes no strict assumptions on data dimensionality, and, lastly, we conclude our results. 

\section{Algorithm}
\subsection{Outline of the Algorithm}
\begin{figure}[ht]
	\centering	
		\includegraphics[width=1\textwidth]{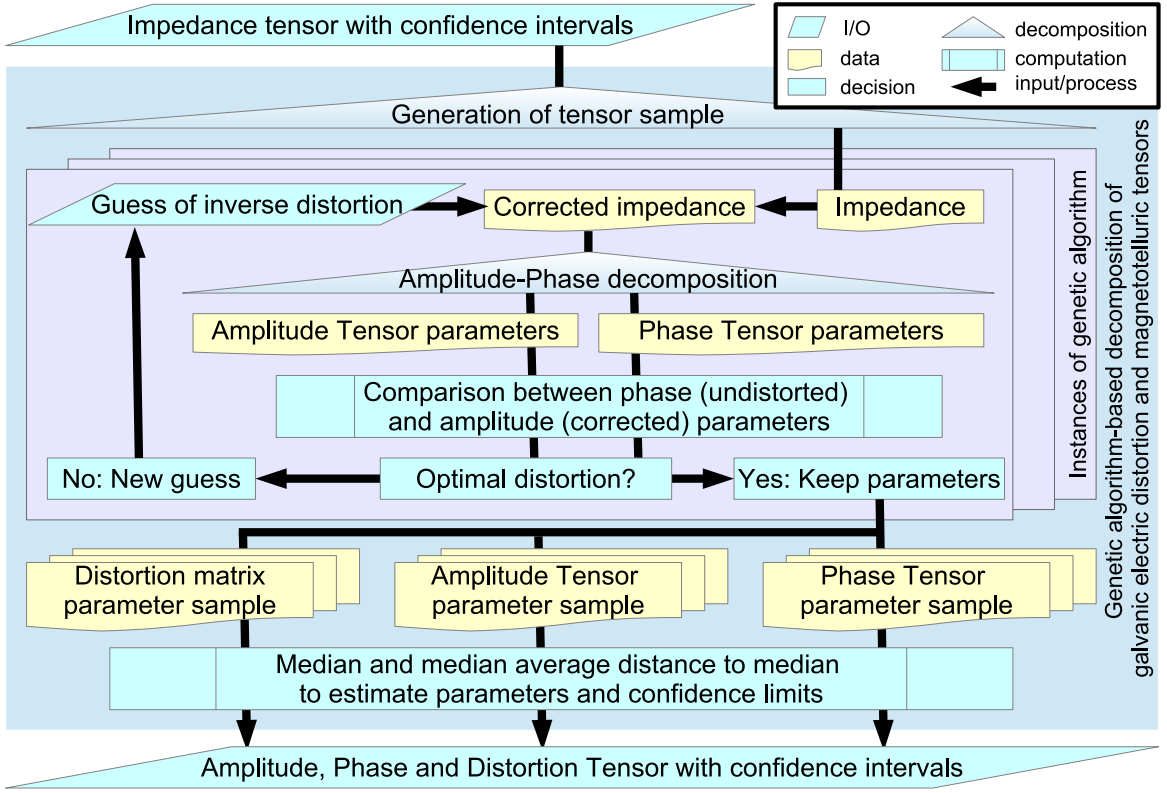}
	\caption{Schematic workflow chart of the proposed procedure to recover the distortion parameters and the regional impedance.}
	\label{fig:workflow}
\end{figure}
Based on the Amplitude-Phase impedance tensor decomposition \citep{Neukirch:2016a} and imposing similarity between inductive AT and PT geometric parameters we have developed an algorithm to estimate the regional MT impedance and the distortion tensor up to the site gain. Figure \ref{fig:workflow} outlines the schematic workflow of this algorithm. 
We refer to the scheme as `Genetic Algorithm-based Decomposition of Galvanic Electric distortion and magnetotelluric Tensors', in short \emph{gadget}. 

Here, we outline the routine and the following sections will describe each step thoroughly. 
\begin{enumerate}
\item Draw an impedance tensor sample from a gaussian distribution with the supplied mean and variance. 
\item Search for the optimal galvanic electric distortion tensor with constant determinant of one, that results in the most similar skew angle, strike angle and anisotropy between AT and PT. A genetic algorithm iterates the following three steps to convergence for each unit of the impedance sample:
\begin{enumerate}
\item Guess distortion matrices $\mathbf{C}$ (population) and attempt to correct the impedance: $\mathbf{Z}=\mathbf{C}^{-1}\mathbf{Z_d}$. 
\item Decompose the corrected impedance into AT and PT parameters \citep{Neukirch:2016a}:
 \begin{equation}
\mathbf{Z} = \mathbf{R}(-\theta_\Phi)  \mathbf{R}(-\gamma)   \mathbf{P_0}  
\mathbf{R}\left(\psi_\Phi+\frac{\pi}{2}+\delta\right)  \mathbf{R}(\gamma)
  \left({c(\mathbf{\Phi_0})} + i  {s(\mathbf{\Phi_0})} 
\mathbf{R}(\psi_\Phi)\right)   \mathbf{R}(\theta_\Phi),
\label{eq:ImpedanceParameter}
\end{equation} 
where $\mathbf{R}$ is the rotation matrix, $\theta_\Phi$ is the phase strike angle, $\gamma=\theta_P-\theta_\Phi$ is the amplitude-phase strike angle difference, $\mathbf{P_0}=\mathrm{diag}(\rho_1,\rho_2)$ contains the amplitude singular values, $\psi_\Phi$ is the phase skew angle, $\delta = \psi_P-\psi_\Phi - \frac{\pi}{2}$ represents the skew angle difference, and ${c(\mathbf{\Phi_0})}=\left(\sqrt{\mathbf{\mathbb{I}}+\mathbf{\Phi_0\Phi_0}^T}\right)^{-1}$ and ${s(\mathbf{\Phi_0})}={c(\mathbf{\Phi_0})\mathbf{\Phi_0}}$ contain with $\mathbf{\Phi_0}=\mathrm{diag}(\phi_1,\phi_2)$ the phase singular values. 

\item Evaluate similarity between AT and PT with respect to skew angle, strike angle and anisotropy using the objective function:
 \begin{equation}
f(\mathbf{C})=\ln{\left|\mathbf{\Psi_P} \mathbf{W_\psi}\mathbf{\Psi_P}^*\right|}+\ln{\left|\mathbf{\Delta} \mathbf{W_\psi}\mathbf{\Delta}^*\right|}+\ln{\left|\mathbf{\Gamma} \mathbf{W_\theta}\mathbf{\Gamma}^*\right|}+{\left|\ln{\left(\mathbf{A_\phi}\mathbf{W_\alpha}\mathbf{A_\phi}^* \right)}-\ln{\left(\mathbf{A_\rho} \mathbf{W_\alpha}\mathbf{A_\rho}^* \right)}\right|},
\label{eq:FitnessFunction}
\end{equation} 
with the diagonal weighting matrices $\mathbf{W}_{\psi/\theta/\alpha}=\mathrm{diag}\left(\frac{f_i^2}{\sigma_{\psi/\theta/\alpha,i}^2\mathbf{f}\mathbf{f}^*}\right)$, associated to the PT skew angle, strike angle and anisotropy variances $\sigma^2_{\psi/\theta/\alpha,i}$, respectively, and vectors of frequency $\mathbf{f}=(f_i)$, and where $\mathbf{\Psi_P}=(\frac{\pi}{2}-\psi_{P,i})$ is the normalised AT skew angle, $\mathbf{\Delta}=(\delta_i)$ is the skew angle difference, $\mathbf{\Gamma}=(\gamma_i)$ is the strike angle difference, $\mathbf{A_\rho}=\left(\alpha_{\rho,i}\right)=\left(\frac{1}{2}\left(\ln{\rho_{1,i}}-\ln{\rho_{2,i}}\right)\right)$ is the logarithmic amplitude anisotropy and $\mathbf{A_\phi}=\left(\alpha_{\phi,i}\right)=\left(\frac{1}{2}\left(\tan^{-1}{\phi_{1,i}}-\tan^{-1}{\phi_{2,i}}\right)\right)$ is the phase anisotropy, where the vector subscript $i$ associates with the frequency. If similarity improves significantly, that is if the objective function decreases from its value in the previous iteration, an educated guess for a new population of distortion matrices (descendants of current population) continues with the search. Otherwise, the genetic algorithm terminates and the best individual is stored in the final parameter samples.
\end{enumerate}
\item Compute robust, non-parametric statistics for the parameter samples resulting from all instances.
\end{enumerate}

\subsection{Step 1: Statistical Considerations and Sample Generation}
 \label{sec:StatisticalConsiderations}
For our statistical analysis of utmost importance is the best possible estimation of confidence intervals, which, if left unconsidered or treated inaccurately, may bias the analysis result and all subsequent interpretation. Additionally, the understanding and analysis of confidence intervals provides insight into systematic problems in data acquisition, processing and analysis schemes alike. 

In our case, although impedance tensor components are usually provided with both mean and variance (typically assuming a gaussian distribution), impedance tensor multiplications imply strong mixing of impedance components and therefore the component variances do not suffice to describe error propagation analytically; the impedance covariances would be required but are not typically supplied. In order to alleviate this problem and to enable the computation of meaningful confidence intervals, we perform the proposed distortion analysis on a large set (typically hundreds) of impedance samples randomly drawn from the assumed gaussian distribution described by the supplied mean and variance. This procedure yields population samples of all analysis results that naturally consider the correlation between impedance components (covariance) while mixing. The final samples are statistically described by robust, non-parametric measures. The median provides the final estimate and the median average distance to the median yields confidence intervals. A parametric approach for the final statistics could result in more realistic location and spread estimates, but for this the parameter distributions must be known or verified for all physically meaningful impedance tensors, which is beyond the scope of this work and therefore, we recourse to the application of non-parametric methods for their applicability to unknown distributions.

Each tensor unit of the generated random sample of the impedance tensor serves as input to an independent instance of the subsequent genetic algorithm. The combination of results from all instances yields statistical samples of all tensor parameters (regional impedance, amplitude, phase and distortion tensors). 

\subsection{Step 2: Genetic Algorithm}
The genetic algorithm is suitable for problems with complex model space landscapes that are typical for multi-objective functions which often encounter many local minima and are prohibitive for direct search algorithms \citep{Gold:1989,Whit:1994}. Alternative approaches could make use of other heuristic techniques, like simulated annealing or a stochastic search like a Markov Chain Monte Carlo algorithm. We chose the genetic algorithm mainly because the problem, along with various remedies for, e.g., getting trapped in a local minima, are very easy to state and convergence to the global minimum is very robust. On the other hand, a major drawback of the genetic algorithm is its long run time due to slow convergence, which, however, is not a major issue for our problem with only three optimisation variables and an objective function that can easily be vectorised for the population. Furthermore, we use a time-tested, sophisticated and efficient genetic algorithm, which is readily available in MatLab (we use version R2014b). 

We set up the genetic algorithm with a population made up of eight subpopulations of various sizes (here adjacent populations with $25$, $75$, $125$, $175$, $225$, $175$, $125$ and $75$ individuals) for which we allow random forward and backward migration ($20\,\%$) between adjacent populations every $20$ iterations. This ensures that there is sufficient time for each subpopulation to find an independent local minimum but also allows vastly superior individuals to spread across several populations to enhance the overall convergence rate. This feature enables eight independent parallel searches which communicate progressively after a consolidation time. Additionally to heuristic breeding of the superior individuals, a crossover rate of $50\,\%$ is applied at each iteration to increase the search radius and to avoid getting trapped in local minima.

Each iteration of the genetic algorithm starts by generating a population of distortion parameter triplets (more details will be given in the following section), first, as a random guess and, subsequently, as an educated guess based on the strongest individuals of the previous generation, until only insignificant improvement is achieved from breeding ($20$ stagnant generations) or an overall limit of iterations is reached ($600$ generations). Usually, the genetic algorithm requires only from $100$ to $300$ iterations to terminate depending on the noise level and/or the severity of distortion present in the data.

To judge the quality of the guess, (i) each distortion tensor individual is inverted and used to correct the original impedance tensor, (ii) the corrected impedance tensor is decomposed into AT and PT parameters, which are (iii) compared for similarity. Similarity of skew angle, strike angle and anisotropy between AT and PT is used as proxy for the quality of the guess of the distortion parameters. The following sections will detail this process, discuss parameter choices for the setting of the genetic algorithm and illuminate the reasoning for the employed objective function.

\subsubsection{Step 2a: Distortion Correction of Impedance Tensor}
Galvanic electric distortion is described by a tensor $\mathbf{C}$ with 4 real-valued components that applies to the regional impedance $\mathbf{Z}$ to yield the distorted, measured impedance $\mathbf{Z}_d=\mathbf{CZ}$ \citep{Bahr:1988,Groom:1989,Chave:1994}. We only consider three variable parameters of the distortion tensor and thus, we assume its determinant, which is also known as the site gain or static shift factor $g$, equal to one. The considered parameters are: twist angle $\phi_t$, shear angle $\phi_s$ and anisotropy angle $\phi_a$, corresponding to the proposed distortion model by \cite{Groom:1989} (see appendix \ref{sec:GalvanicDistortion} for details). We chose this model because the parameters are physically intuitive and, especially, the twist angle displays a very clear distortion signature on amplitude skew and strike angles facilitating the comparison of similarity between the amplitude and the phase (see appendix \ref{sec:GalvanicDistortion} for more detail). The distortion tensor determinant is chosen to be fixed at the value of one, because, to the knowledge of the authors, there is no method currently known that can estimate this parameter from the data of a single site without information provided by other electromagnetic data. 


If the distortion tensor is known, the impedance can be corrected by 
 \begin{equation}
\mathbf{Z}= (\mathbf{C}(\phi_t,\phi_s,\phi_a))^{-1}   \mathbf{Z_d}.
\label{eq:CorrectImpedance}
\end{equation} 
We use this equation with each individual of the population of distortion parameter triplets to compute \emph{corrected} impedance tensors, which are assessed for their similarity between their AT and PT. 
In order to evaluate similarity, the tensor parameters are compared and the following section elucidates the employed tensor decomposition and parameterisation.

\subsubsection{Step 2b: Amplitude-Phase Tensor Decomposition and Parameterisation}
The MT impedance tensor can be completely decomposed into the multiplication of the real-valued AT and a function of the real-valued PT as shown in \eqref{eq:AmplitudePhaseDecomposition}. The phase function $e(\mathbf{\Phi})$ \citep{Neukirch:2016a} of the PT $\mathbf{\Phi}$ \citep{Caldwell:2004} yields with \eqref{eq:AmplitudePhaseDecomposition} the AT: $\mathbf{P} =\mathbf{Z}  \left(e(\mathbf{\Phi})\right)^{-1}$.
Then, both tensors can be parameterised following \cite{Booker:2014} for $\mathbf{M}\in\mathbb{R}^{2\times2}$:
 \begin{equation}
\mathbf{M} = \begin{pmatrix} M_{1,1} & M_{1,2}\\M_{2,1} & M_{2,2} \end{pmatrix} = \mathbf{R}(-\theta_M)   \begin{pmatrix} m_1 & 0\\0 & m_2 \end{pmatrix}  \mathbf{R}(\psi_M)   \mathbf{R}(\theta_M), 
\label{eq:TensorParameter}
\end{equation} 
where $\theta_M$ is a coordinate rotation angle, describing the angle between the matrix coordinate system and the cartesian coordiante system, 
 \begin{equation}
	\label{psi_M}
  \psi_M=\begin{cases}
    \arctan\frac{M_{1,2}-M_{2,1}}{M_{1,1}+M_{2,2}}, & \text{if $0\le|M_{1,2}-M_{2,1}| \le |M_{1,1}+M_{2,2}| \ne 0$},\\
    \arccot\frac{M_{1,1}+M_{2,2}}{M_{1,2}-M_{2,1}}, & \text{if $|M_{1,2}-M_{2,1}| > |M_{1,1}+M_{2,2}|$ and $\frac{M_{1,1}+M_{2,2}}{M_{1,2}-M_{2,1}}\ge0$},\\
    \arccot\frac{M_{1,1}+M_{2,2}}{M_{1,2}-M_{2,1}}-\pi, & \text{if $|M_{1,2}-M_{2,1}| > |M_{1,1}+M_{2,2}|$ and $\frac{M_{1,1}+M_{2,2}}{M_{1,2}-M_{2,1}}<0$},
  \end{cases}
\end{equation} 
is the normalised matrix skew angle \citep{Neukirch:2016a}, $m_1$ and $m_2$ are the singular values, and $\mathbf{R}(\xi)$ is the (orthogonal) rotation matrix with the rotation angle $\xi$:
 \begin{equation}
\mathbf{R} (\xi) = \begin{pmatrix} \cos(\xi) & \sin(\xi)\\-\sin(\xi) & \cos(\xi) \end{pmatrix}.
\label{eq:RotMat}
\end{equation} 
This parameterisation can be brought further into a form that contains one scale factor (determinant) and three geometric parameters (skew angle, strike angle and anisotropy) that describe the subsurface geometry and dimensionality. 

The choice of the decomposition and parameterisation above are motivated by two important considerations of our proposed approach to recover the galvanic distortion parameters, that is (i) physical similarity and (ii) parameter comparability of Amplitude and Phase Tensor parameters. 

The first consideration is the reason for the principal choice of employing the Amplitude-Phase decomposition and is based on the similarity hypothesis by \cite{Neukirch:2016a}. 
Additionally, the tensor parameters skew and strike angle, each describe physically intuitive geometric and dimensional features that must be related to the subsurface, offering an ideal base for comparing the two tensors on these terms. The second consideration is automatically ensured for angular parameters like skew and strike angles, demonstrating the reason for the choice of the parameterisation. Another geometric parameter is amplitude and phase anisotropy which is defined by the discrepancy between the respective singular values. As pointed out by \cite{Neukirch:2016a} the amplitude and phase anisotropy cannot be compared directly on their default linear scale but a logarithmic scale, which relates phase and logarithmic amplitude like real and imaginary parts of the natural logarithm of the 1D impedance:
 \begin{equation}
Z_{1D}=|Z_{1D}|\exp{\left(i\phi_{1D}\right)}=\exp{\left(\ln|Z_{1D}|+i\phi_{1D}\right)}=\exp{\left(\ln Z_{1D}\right)}.
\end{equation} 
In order to compare amplitude and phase anisotropy, we denote the phase (amplitude) singular values as $\phi_1$ ($\rho_1$) and $\phi_2$ ($\rho_2$) and we consider the following scale factors:
 \begin{equation}
\phi_0=\frac{1}{2}\left(\tan^{-1}(\phi_1)+\tan^{-1}(\phi_2)\right)\qquad \mathrm{and}\qquad
\rho_0 = \frac{1}{2}\left(\ln(\rho_1)+\ln(\rho_2)\right)
\label{eq:Scales}
\end{equation} 
and the phase anisotropy and logarithmic amplitude anisotropy \citep{Neukirch:2016a}:
 \begin{equation}
\alpha_\phi =\frac{1}{2}\left(\tan^{-1}(\phi_1)-\tan^{-1}(\phi_2)\right)\qquad \mathrm{and}\qquad
\alpha_\rho = \frac{1}{2}\left(\ln(\rho_1)-\ln(\rho_2)\right).
\label{eq:Anisotropies}
\end{equation} 
Appendix \ref{sec:GalvanicDistortion} continues to discuss parameter similarity and comparability on an explicit example.

Finally, from the decomposition in \eqref{eq:AmplitudePhaseDecomposition} and the parameterisation in \eqref{eq:TensorParameter}, we obtain the parameters for the impedance tensor, $\mathbf{Z}$, as is given in \eqref{eq:ImpedanceParameter}.

The following section will illustrate how the numerical comparison of amplitude and phase skew angle, strike angle and anisotropy is brought into the form of an objective function, which is used to minimise the  dissimilarity between the parameters, imposing the physical considerations of an unique subsurface geometry and dimensionality.

\subsubsection{Step 2c: Comparison of Skew, Strike and Anisotropy}
\label{sec:ComparisonFunction}
\cite{Neukirch:2016a} postulated that for a purely inductive response of the subsurface, AT and PT skew angle, strike angle and anisotropy must be similar. This means in particular, that all three are similar when all galvanic effects are stripped from the impedance at a certain period 
as in \eqref{eq:CorrectImpedance}. To quantify similarity, we define the objective function \eqref{eq:FitnessFunction}, considering \eqref{eq:ImpedanceParameter}, \eqref{eq:CorrectImpedance} and \eqref{eq:Anisotropies} for each distortion parameter triplet, $\phi_t$, $\phi_s$ and $\phi_a$.

The objective function \eqref{eq:FitnessFunction} is a multi-objective function consisting of four sub-objectives related to the AT skew angle, the Amplitude-Phase Tensor skew angle difference, the Amplitude-Phase Tensor strike angle difference and the difference of the logarithm of AT and PT anisotropy parameters. The first part is a stability term that increases the convergence rate by penalising large distortion parameters and homogenising short period data. This term originates from the assumption that 3D effects in the AT at short periods are attempted to be described by galvanic distortion of the electric field as much as possible but within the constraints of the following sub-objectives. The latter three parts prioritise Amplitude-Phase Tensor similarity between the respective tensor parameters. The sub-objective functions are expressed on logarithmic scale, because of faster convergence rates and because the similarity of anisotropy is best described logarithmically as discussed in the previous section. We refer to appendix \ref{sec:GalvanicDistortion} for a scrutiny of distortion effects on the AT in contrast to the PT and the reasoning behind the hypothesis of Amplitude-Phase Tensor similarity for regional inductive responses.

The weighting function is chosen to be dimensionless when multiplied with the data and to give preference to low variance and short period data, such that the distortion correction is attempted for the superficial part of the data that has a more-than-average confidence. Since the PT is undistorted, the variances for $\mathbf{\Psi_P}$, $\mathbf{\Delta}$, $\mathbf{\Gamma}$ and $\mathbf{A_\rho}$ are estimated from the corresponding PT parameter variances, i.e.~the variance of $\psi_\Phi$, $\theta_\Phi$ and $\alpha_\phi$, which are known a-priori to the execution of the genetic algorithm. This estimation of variances is justified because associated AT and PT parameters appear to produce similar variance characteristics \citep{Neukirch:2016a}. 
For more details on the choice of the weighting matrices, in particular the inter-objective weighting, we refer to appendix \ref{ConsiderationsObjectiveFunction}.

The objective function evaluates each triplet of distortion parameters, $\phi_t$, $\phi_s$ and $\phi_a$, and provides a mean to quantify similarity of AT and PT and therewith their appropriateness to correct the distorted impedance. In every iteration, the best triplets (individuals) of the population are chosen to breed a new generation for the next search iteration until the objective function changes are stagnant for $20$ generations and the genetic algorithm terminates. In addition to breeding between the best individuals, random mutations increase the search radius and allow the algorithm to explore solutions beyond local minima. Furthermore the division of the population into several subpopulations along with random migration every $20$ iterations allows to approach several local minima in parallel, enhances the effect of mutation and improves convergence rates due to inherent parallelism. 

Once the genetic algorithm terminates, the best individual along with the associated tensor parameters are stored in the parameter samples together with the results of all other instances of the genetic algorithm (of which each corresponds to a drawn impedance from the gaussian impedance distribution as discussed in section \ref{sec:StatisticalConsiderations}). This procedure yields a parameter sample for the distortion parameters $\phi_t$, $\phi_s$ and $\phi_a$, the undistorted AT and PT parameters at each period $\psi_{P/\Phi}$, $\theta_{P/\Phi}$, $\rho_{1,2}$ and $\phi_{1,2}$, and the undistorted, complex-valued impedance components, $Z_{xx}$, $Z_{xy}$, $Z_{yx}$ and $Z_{yy}$, at each period. The following section will lay out the statistical analysis of these final parameter samples.


\subsection{Step 3: Statistics of the Results}
Since the Amplitude-Phase decomposition and parameterisation is non-linear, we cannot assume that a gaussian distribution, as usually assumed for the original impedance measurements, is suitable to describe all parameter samples. In fact, it is unlikely that any of the parameter samples follows a gaussian distribution. 
Instead, we recourse to non-parametric measures which are robust and meaningful on a wide range of possible distributions and provide in most cases a reasonable good estimate of location and spread. We advocate the median to determine the location and the median average distance to the median (MAD) for the spread.

Although, we do not assume any particular distribution for any parameter sample, it is clear that some are based on particular ranges, with the angular parameters being pertinent examples that often are defined on $[0\ge \beta>2\pi]$ with both limits adjacent to each other and that require circular statistical treatment \citep{fisher:1995,jammalamadaka:2001,zar:1999}. In our case, we distinguish three classes of parameters related to their scale: linear, logarithmic and angular, and treat each class differently. Linear and logarithmic parameters are based on a linear and a logarithmic scale, respectively, and for angular parameters, we suggest to define distinct ranges for each parameter appropriate to the information described by that parameter. We assume as linear parameters all eight impedance components (real and imaginary parts) and the PT singular values. Logarithmic parameters are the two singular values of the AT. The rest are angular parameters. The distortion parameters twist, shear and anisotropy angles are defined on a range of $[-\frac{\pi}{2}< \phi_t<\frac{\pi}{2}]$, $[-\frac{\pi}{4}< \phi_t<\frac{\pi}{4}]$ and $[-\frac{\pi}{4}< \phi_a<\frac{\pi}{4}]$, respectively, to ensure a unique decomposition of the distortion tensor \citep{Groom:1989}. The skew and strike of AT and PT assume the range $[-\frac{\pi}{2} < \psi \le \frac{\pi}{2}]$ and $[0 \le \theta < \frac{\pi}{2}]$, respectively, where tangent periodicity and strike angle ambiguity has been considered. 

\section{Applications}
\subsection{Test Data Set: Dublin Secret Model 2}
\label{sec:EM3DII}
\begin{figure}[htp]
	\centering
		\includegraphics[width=.31\textwidth]{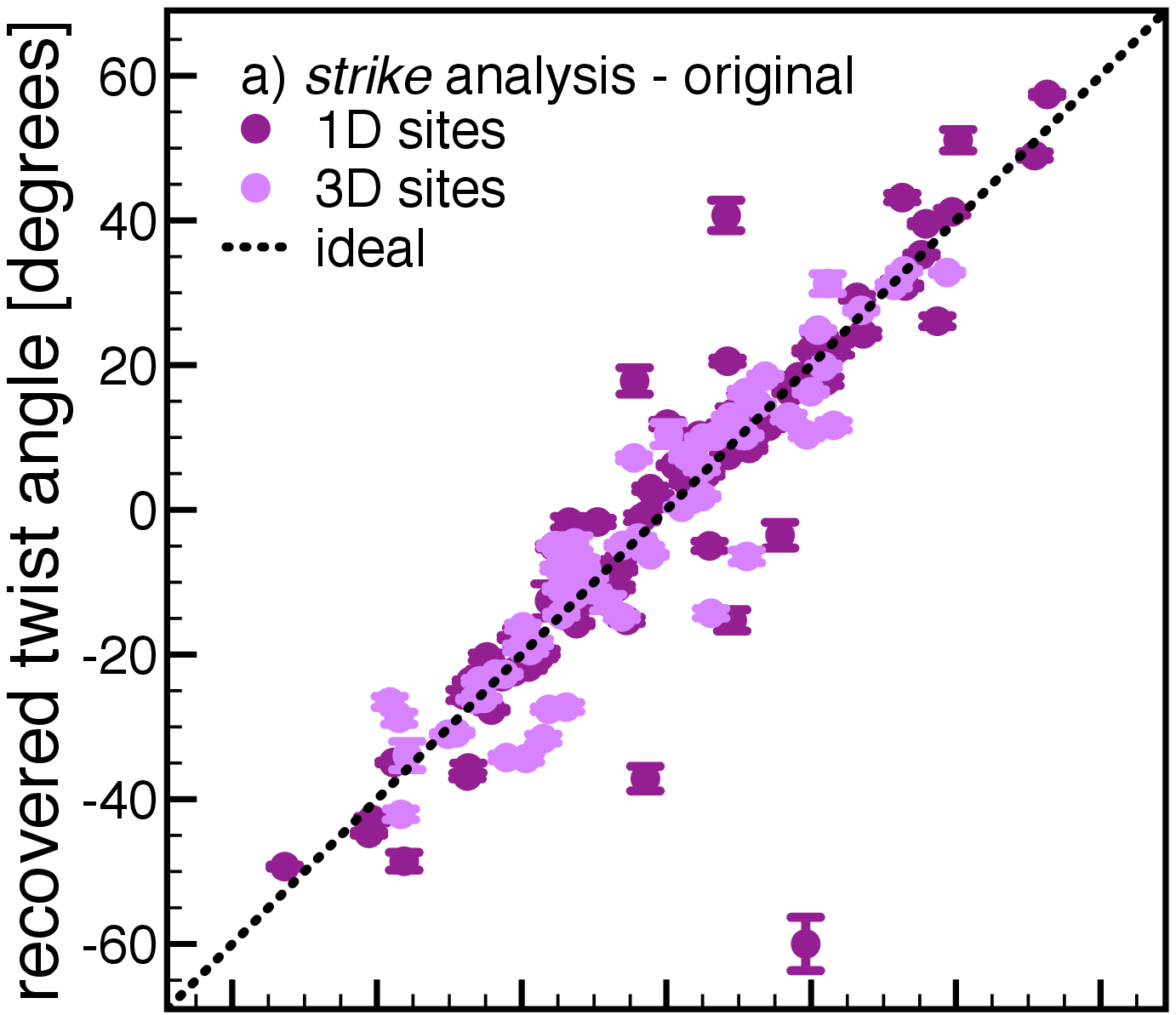}
		\includegraphics[width=.31\textwidth]{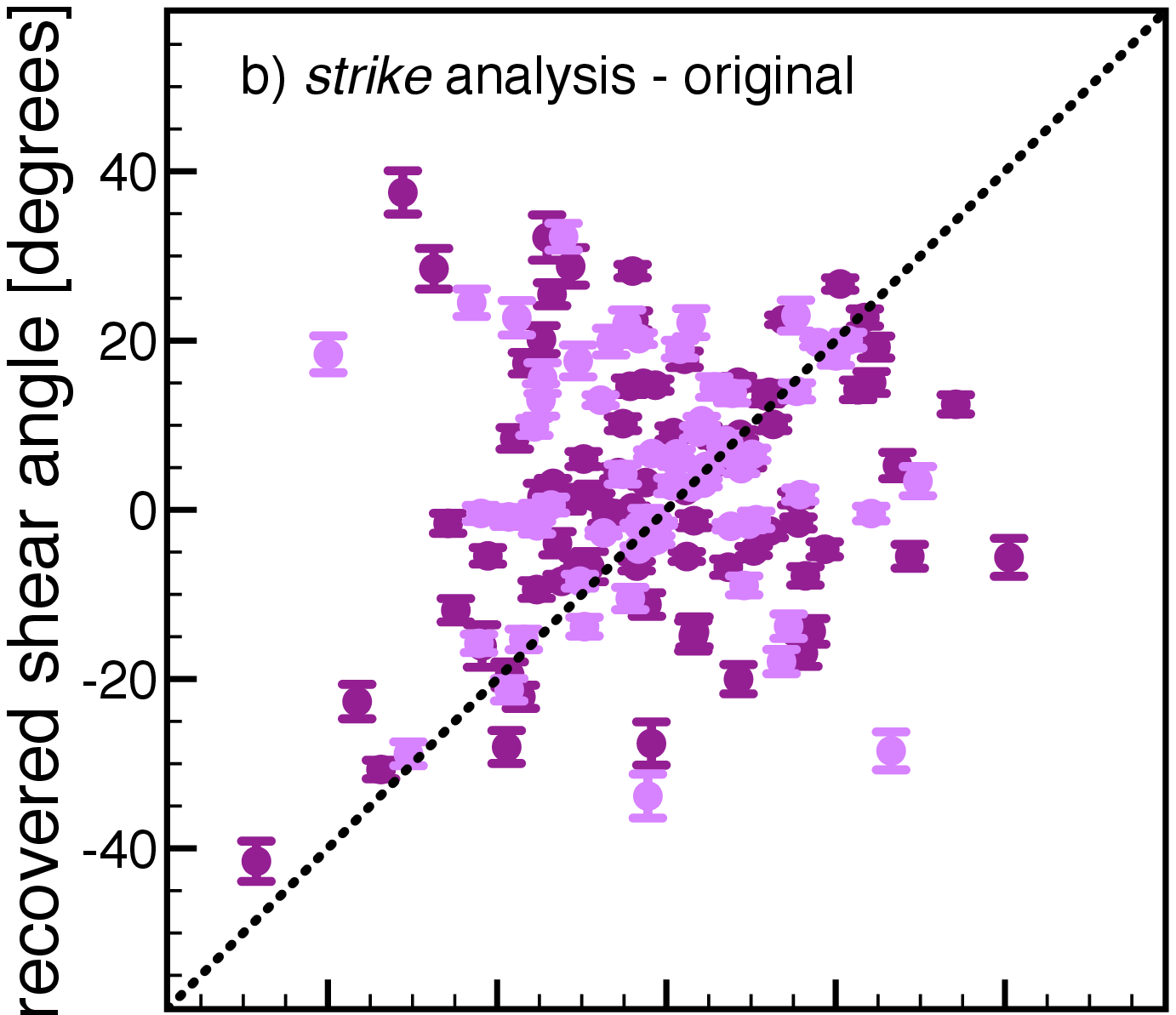}
		\includegraphics[width=.31\textwidth]{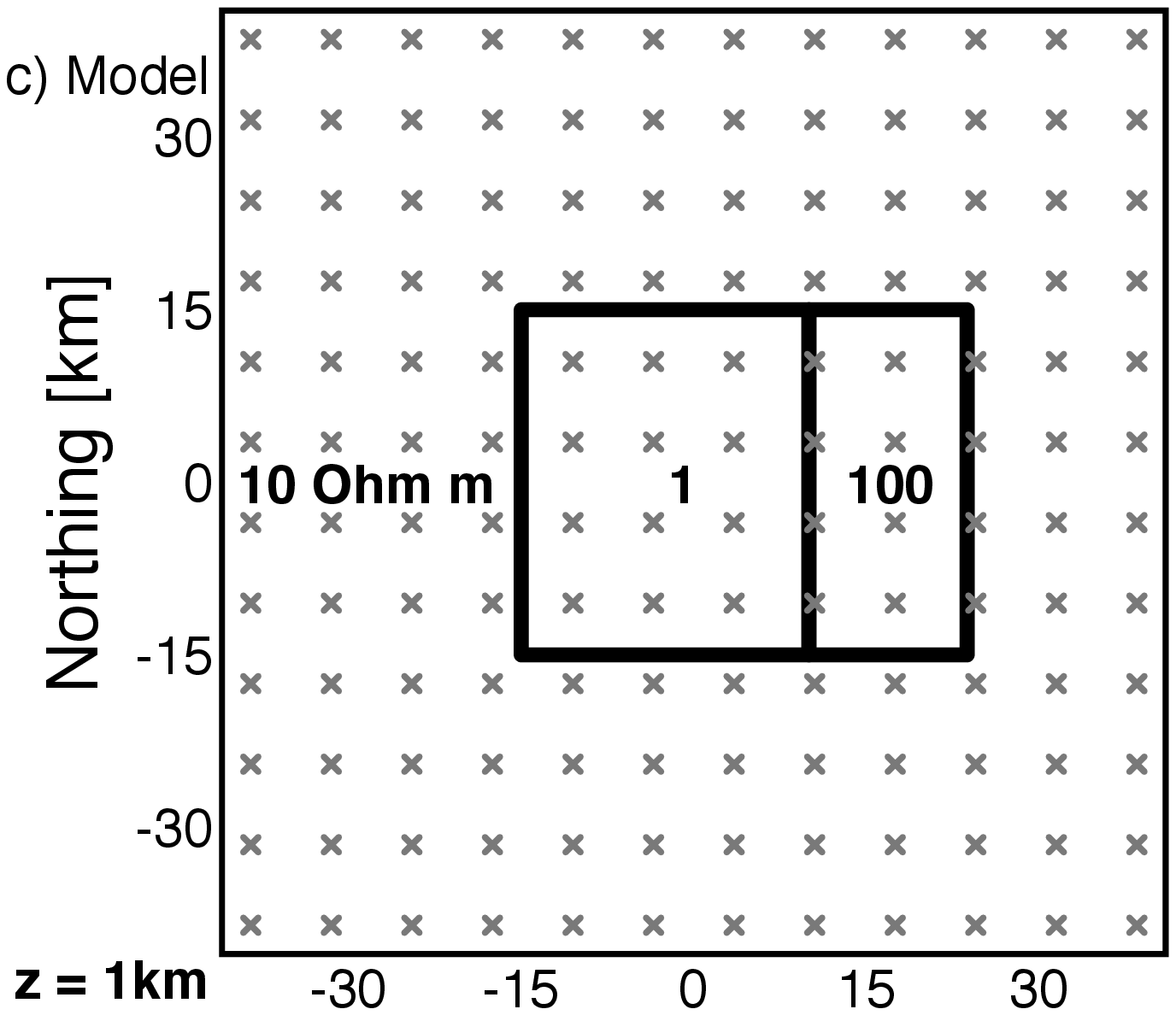}\\
		\includegraphics[width=.31\textwidth]{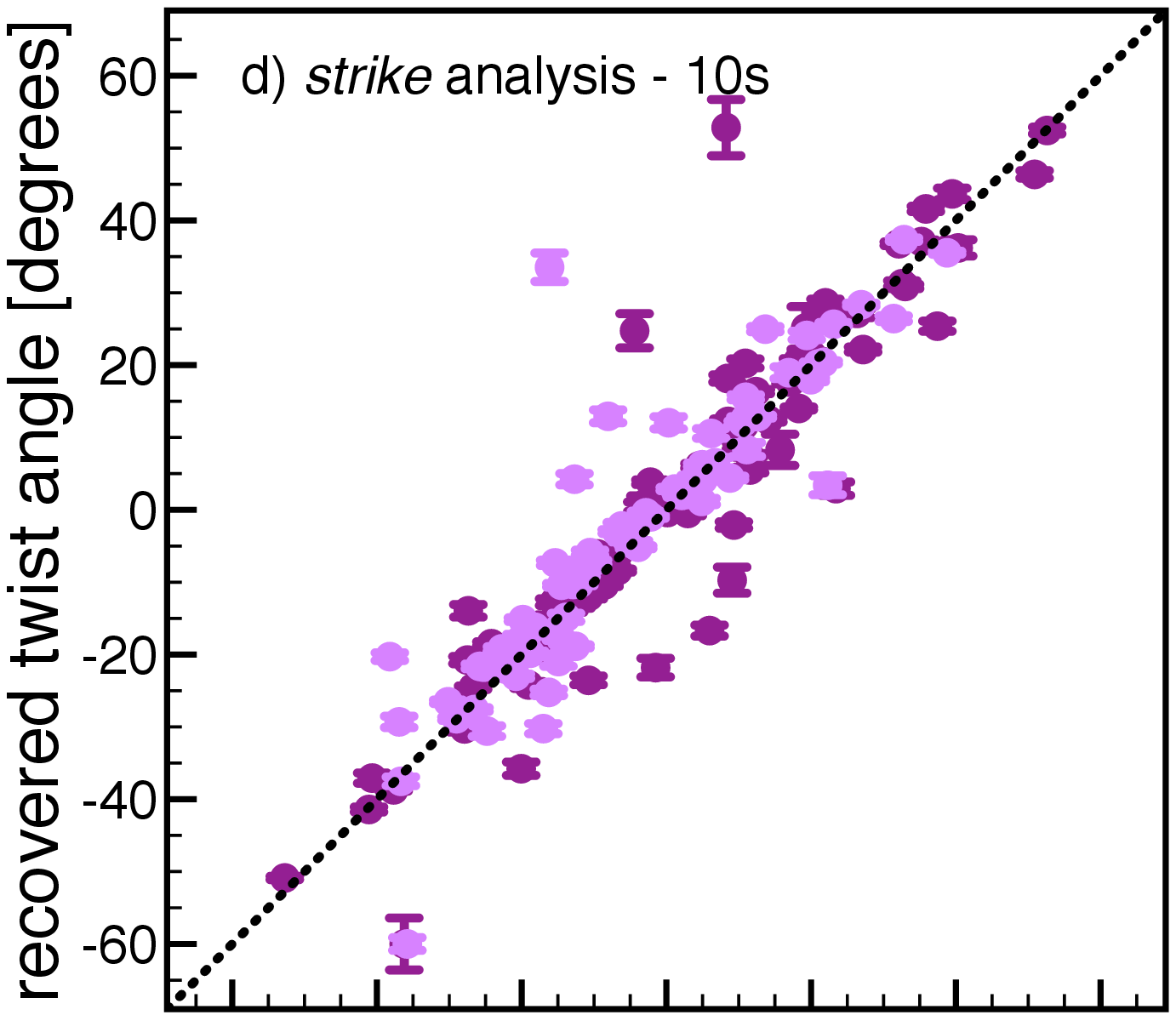}
		\includegraphics[width=.31\textwidth]{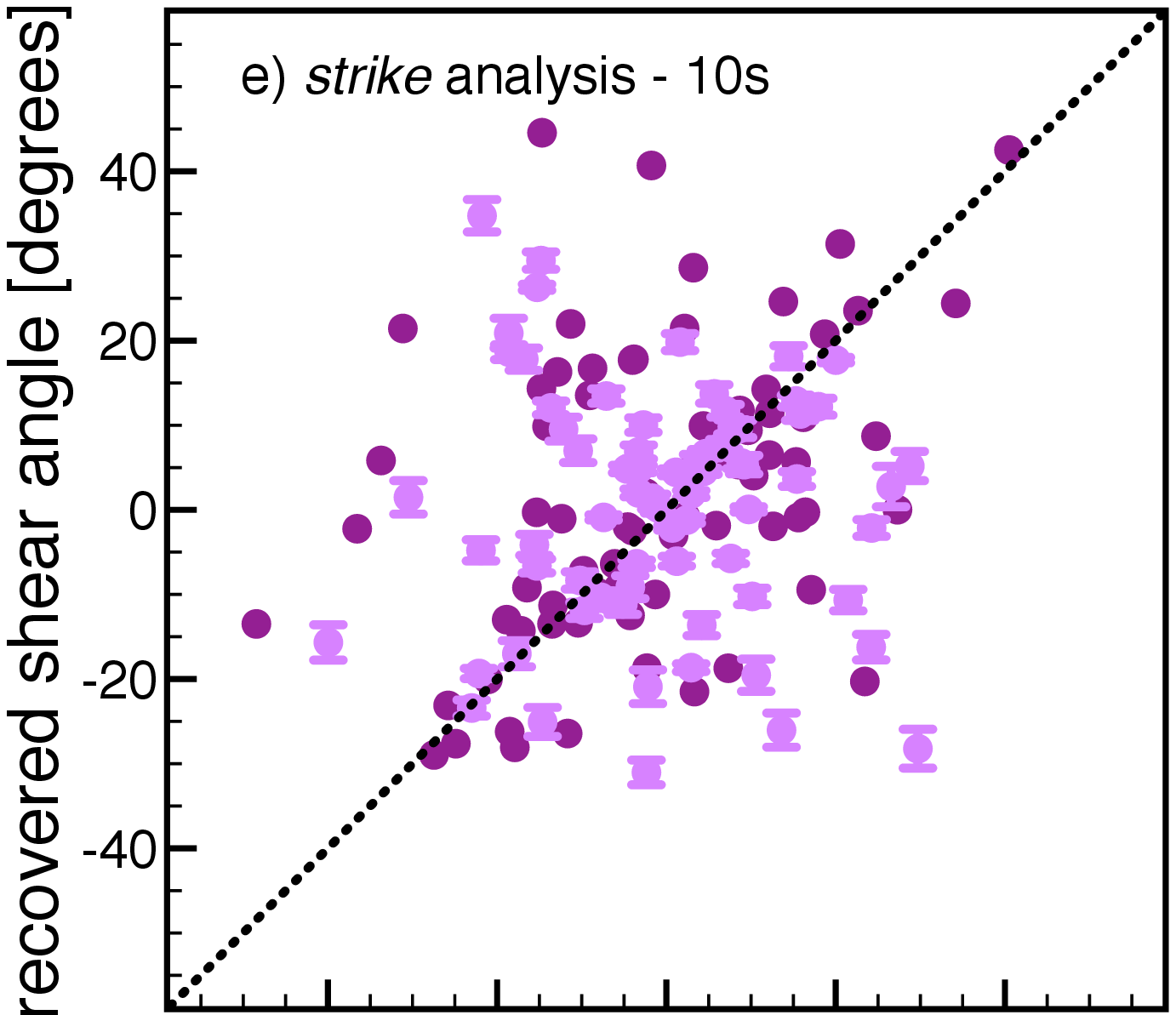}
		\includegraphics[width=.31\textwidth]{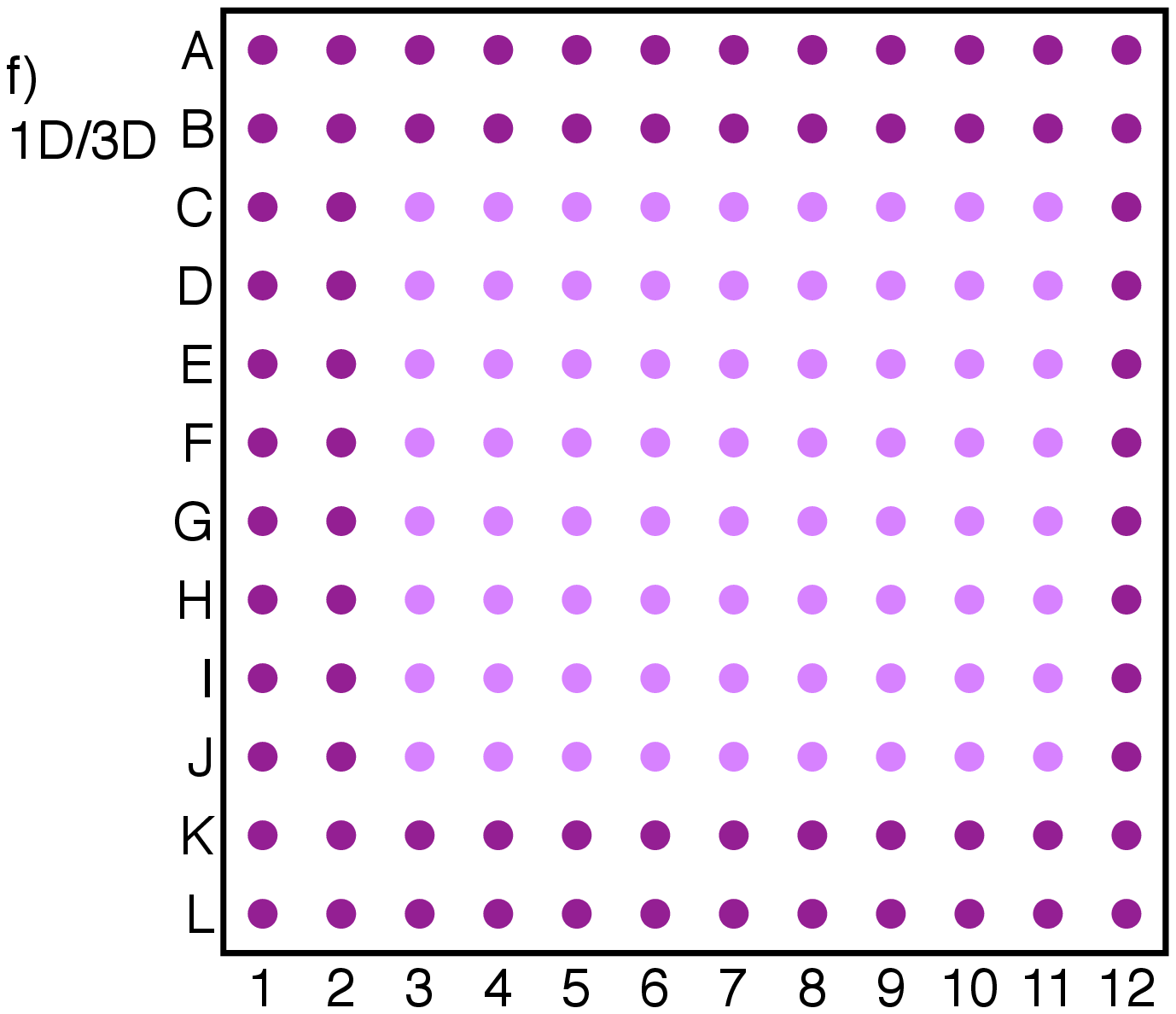}\\
		\includegraphics[width=.31\textwidth]{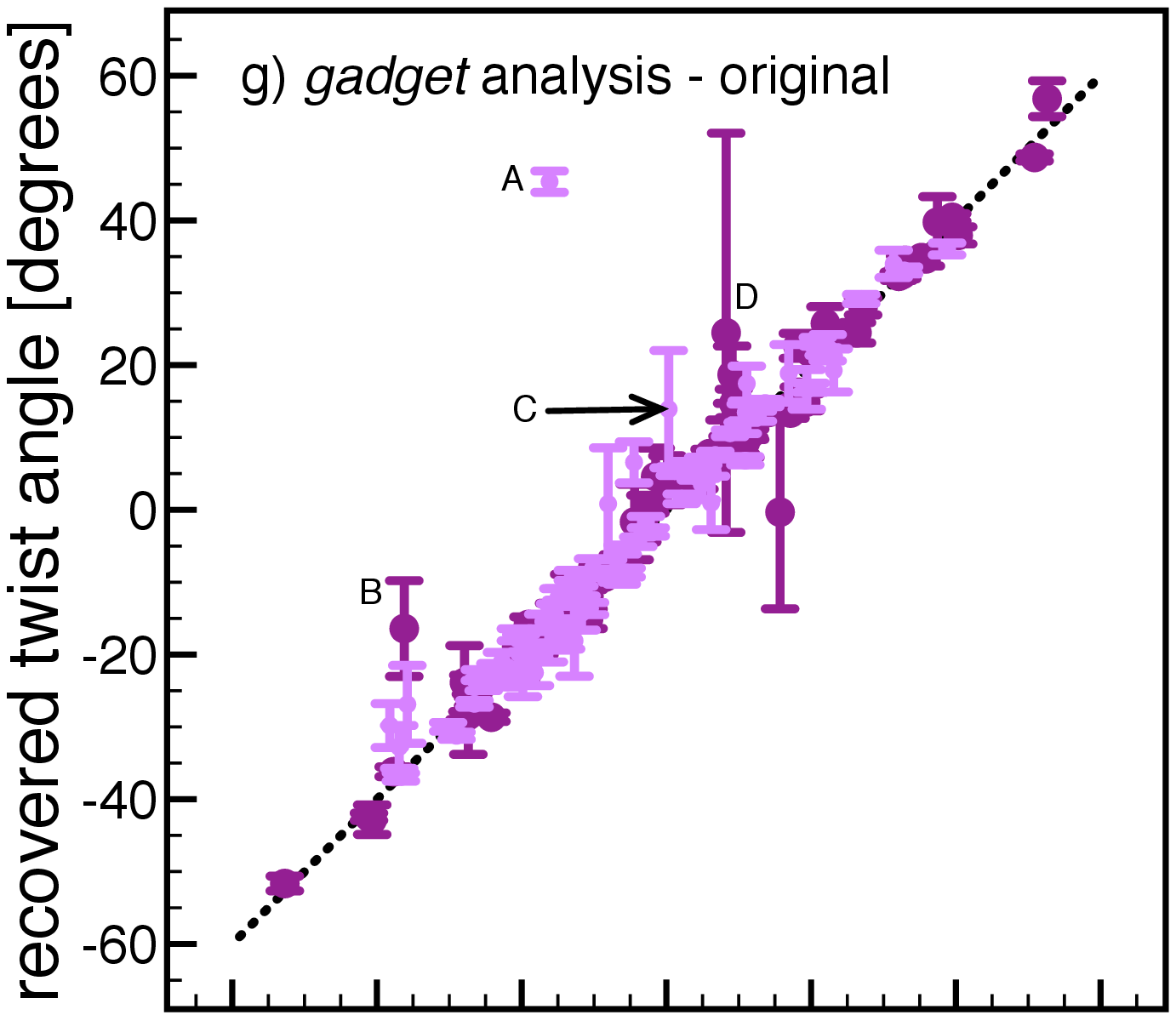}
		\includegraphics[width=.31\textwidth]{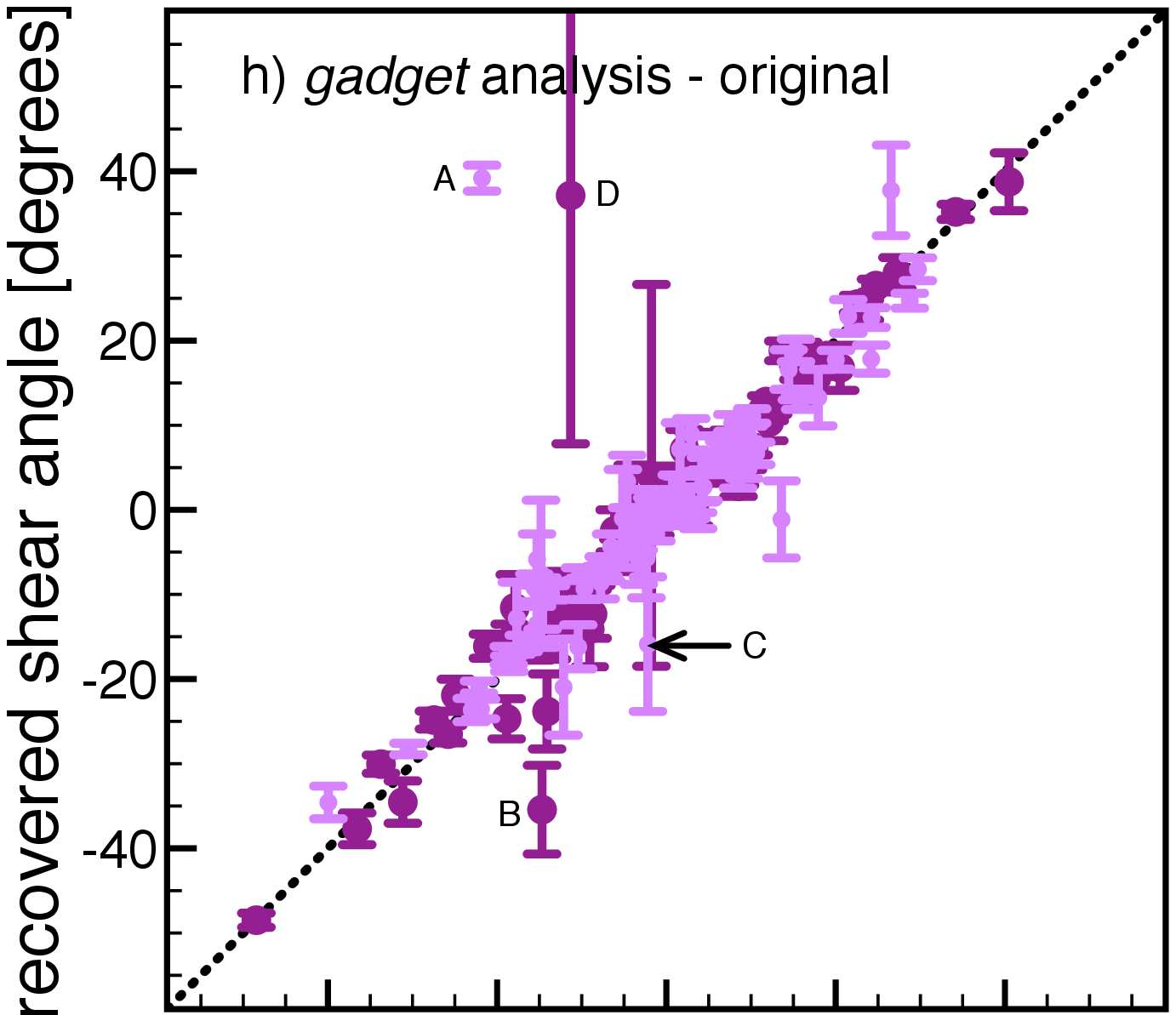}
		\includegraphics[width=.31\textwidth]{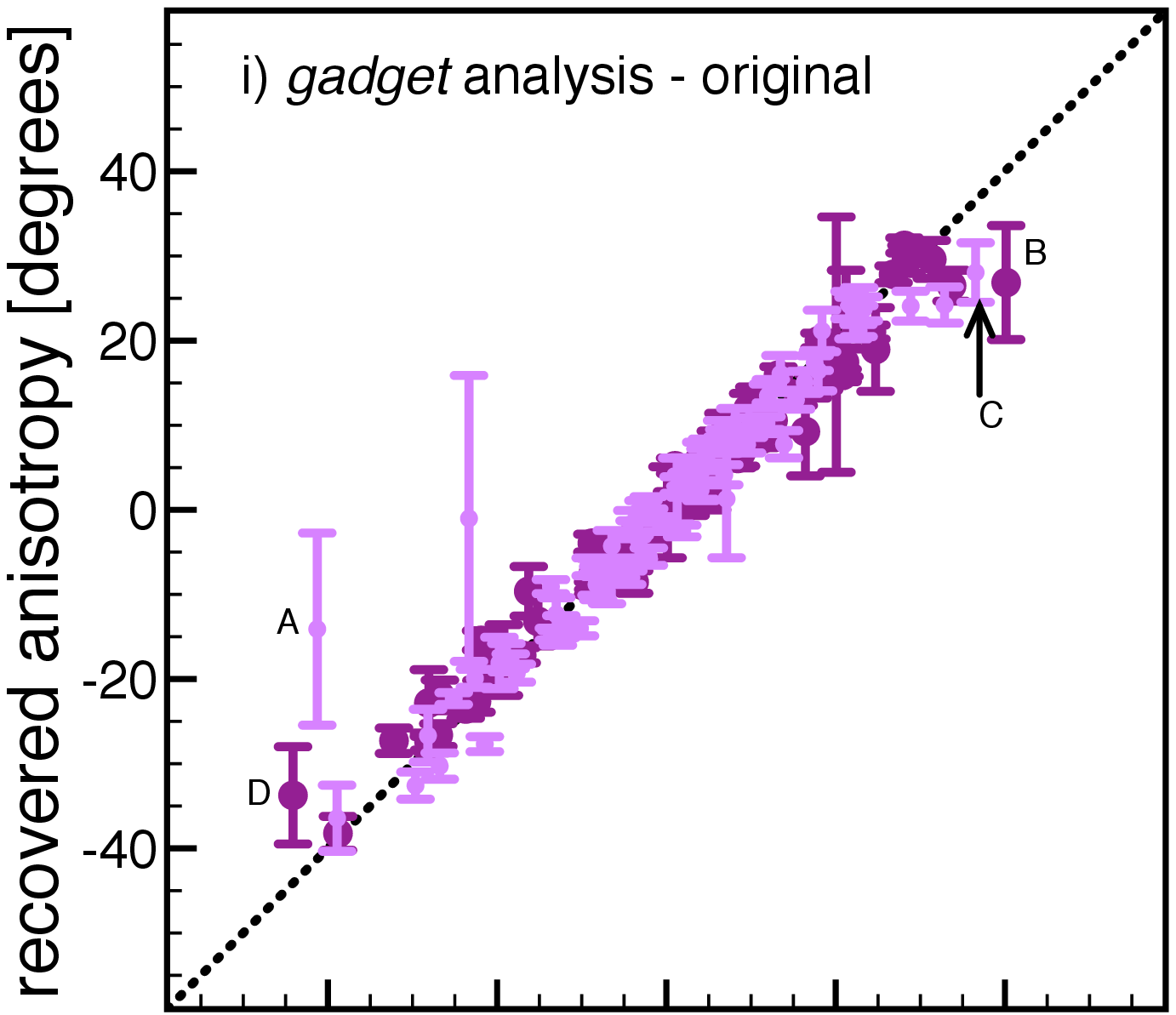}\\
		\includegraphics[width=.31\textwidth]{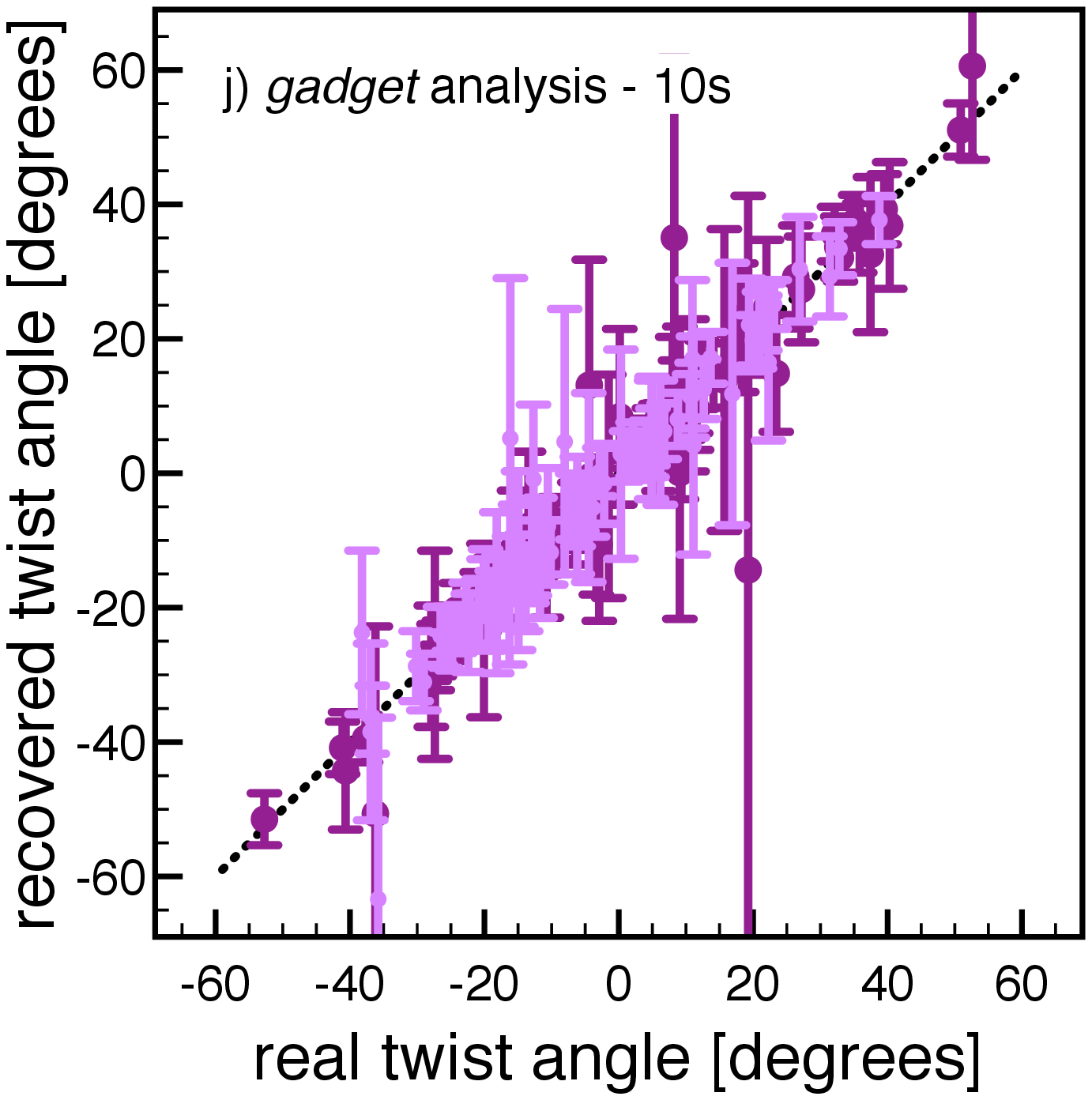}
		\includegraphics[width=.31\textwidth]{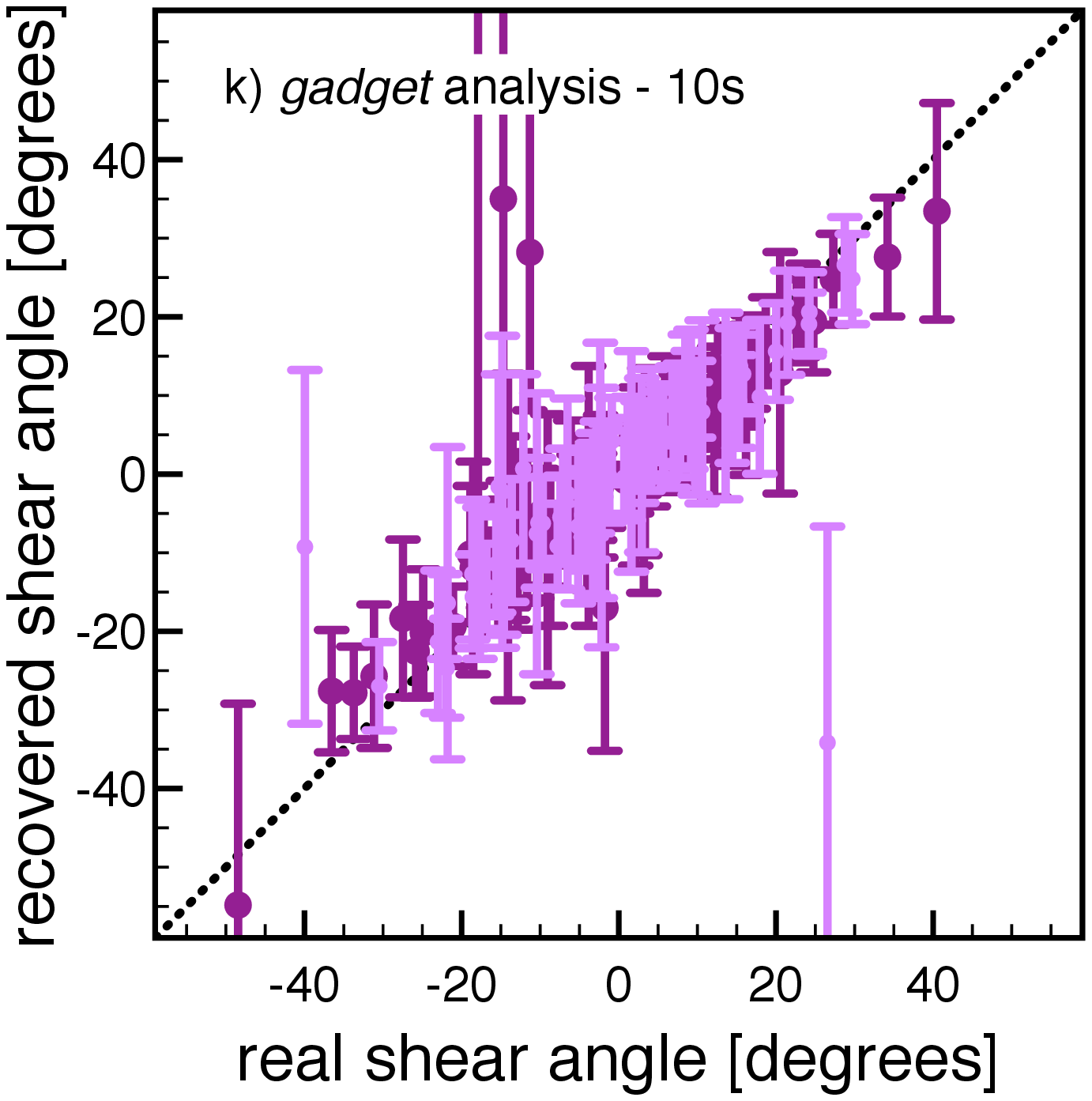}
		\includegraphics[width=.31\textwidth]{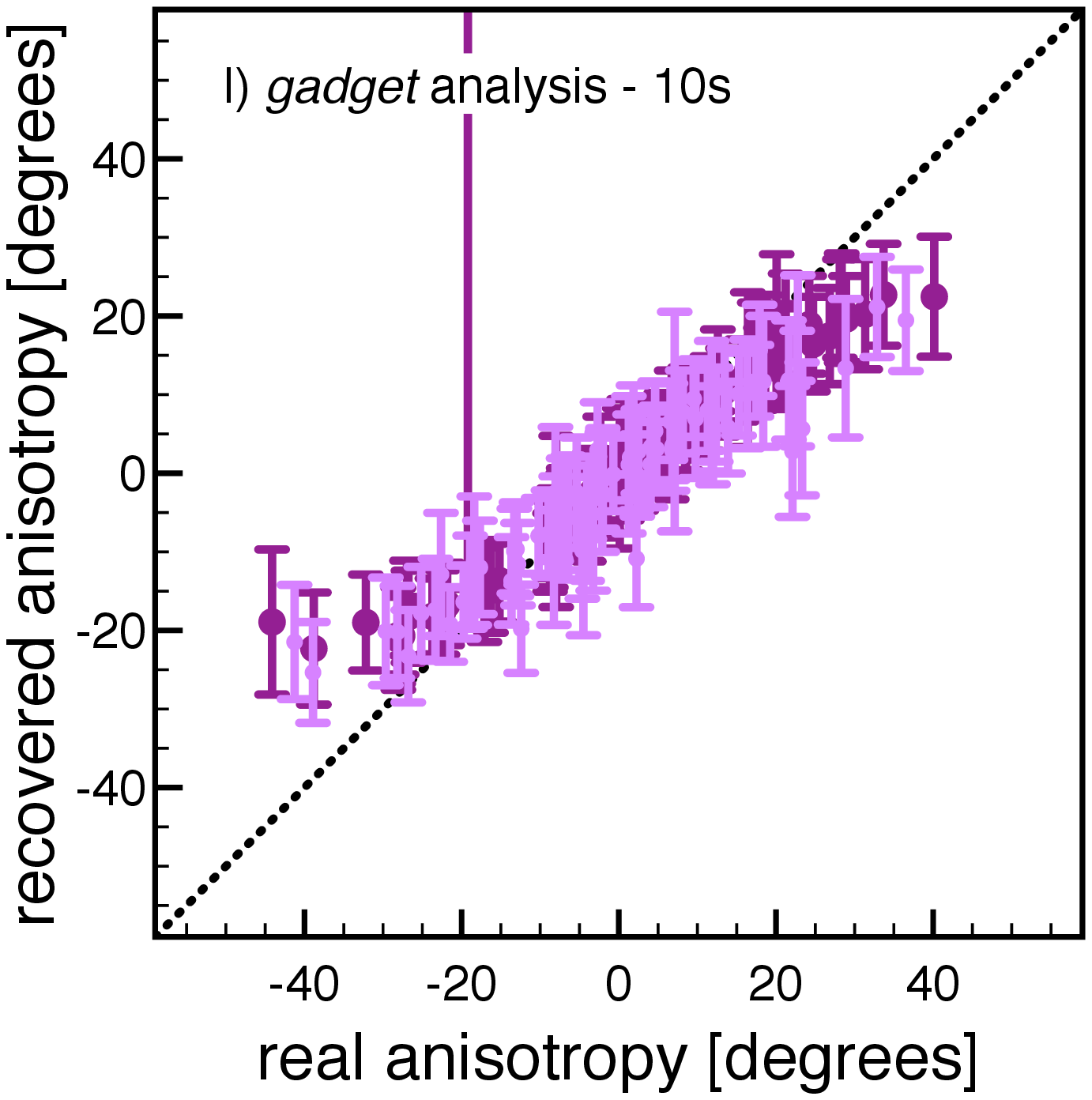}\\
	\caption{Original and recovered twist, shear and anisotropy angles from the DSM2 data set \citep{Miensopust:2013} are plotted against each other and using two distortion analysis programs: \emph{strike} and \emph{gadget}. A bird's eye view of the model and the site locations are plotted (c and f). To demonstrate the performance for 3D situations, a second run omitted period data of less than $10\,\mathrm{s}$ so that the shortest period data (skin depth of $11\,\mathrm{km}$ for $50\,\Omega m$) penetrate the homogenous upper layer ($1\,\mathrm{km}$ of $50\,\mathrm{\Omega m}$) and sense 3D features at the indicated sites. \emph{Strike} yields moderately good estimates for twist but not for shear with little difference with respect to the choice of 1D or 3D data and is more likely controlled by the high noise level (which \emph{strike} does not reflect in the confidence intervals). Despite the presence of noise, \emph{gadget} delivers reliable estimates for nearly all parameters, demonstrating noise resistance and suitability for 3D data. The most prominent outliers are labelled (A, B, C and D) in the original data to emphasise that high anisotropic distortion ($|\phi_a|>25^{\circ}$) impedes the algorithm, whereas high shear or twist can be recovered successfully.} 
	\label{fig:DSM2}
\end{figure}
Before turning to real data, let us demonstrate the performance of the proposed
methodology on a synthetic data set from the 3D Modelling and Inversion
Workshop held in Dublin in 2011, the Dublin Secret Model Two \citep[DSM2,
][]{Miensopust:2013}. The model contains two $10\,\mathrm{km}$ deep blocks of $1\,\mathrm{\Omega m}$ and $100\,\mathrm{\Omega m}$ in a background resistivity of $10\,\mathrm{\Omega m}$, all covered by a $1\,\mathrm{km}$ thick layer of $50\,\mathrm{\Omega m}$ (cp.~Figure~\ref{fig:DSM2}c). 
The interesting feature of this synthetic 3D data set is
that the modelled forward responses have been modified by adding random distortion and
noise in order to resemble a real data acquisition and test how well inversion
codes cope with those `realistic' problems, i.e.~by \cite{avdeeva:2015}. 

We perform a distortion analysis on the entire data set using \emph{gadget} and another program, \emph{strike} \citep{McNeice:2001}, in order to benchmark our program. \emph{Strike} is a powerful algorithm that statistically estimates distortion and geometrical parameters from the data by assuming a regional electric 2D subsurface with 3D superficial distortion. The algorithm can be used to estimate these parameters on a frequency per frequency basis \citep[similar to][]{Groom:1989} but the real power of the algorithm lies in its established statistical framework for a multi-frequency and multi-site analysis. Since the present data set is based on a 3D model and the inclusion of multiple sites and a large range of frequencies breaks the assumption of 2D, we only perform the \emph{strike} analysis for the shortest period and at each site separately. The shortest period of the data set is $T_{\mathrm{min}}=0.158\,\mathrm{s}$ and corresponds to a skin depth of $\delta_{\mathrm{min}}=448\,\mathrm{m}$ that is well within the superficial homogenous layer and thus allows us to treat the shortest period data as 1D. 

Figures~\ref{fig:DSM2}a and \ref{fig:DSM2}b display the \emph{strike} results for the shortest period data as cross plots of the predicted versus the real distortion parameters. The twist distortion angle is recovered for much of the data to some precision, but the estimation of the accuracy does not represent the present noise and data uncertainty. The shear angle estimate appears to be very inaccurate without a reliable estimate of data confidence. 

In order to also test performance for real 3D data (at the analysed period), we perform a second analysis omitting all data at periods below $10\,\mathrm{s}$. Then, the shortest period data corresponds to the skin depth $\delta=11.25\,\mathrm{km}$ for a $50\,\mathrm{\Omega m}$ half-space and it is ensured that the signal has penetrated the upper layer and adjacent site's sensitivity overlaps on 3D model features. A site map showing which data must penetrate into 3D features for $T_{\mathrm{min}}=10\,\mathrm{s}$ is illustrated in Figure~\ref{fig:DSM2}f. Figures~\ref{fig:DSM2}d and \ref{fig:DSM2}e display the \emph{strike} results for this data as cross plots of the predicted versus the real distortion parameters. There is no notable difference between the analysis results of the $T_{1}=0.158\,\mathrm{s}$ and $T_{2}=10\,\mathrm{s}$ data, and our interpretation is that the results are noise driven rather than due to subsurface dimensionality. This interpretation is based on the large scatter of shear angle results observed in Figure \ref{fig:DSM2}b for the data at $T_1$. The data should be clearly within the regional 2D assumption of \emph{strike}, but the algorithm is not able to estimate the shear angles accurately and underestimates the confidence limits, reflecting, probably, the presence of strong noise in the impedance data. In contrast, it appears that the twist angle estimation is more robust against the presence of noise and does not necessarily require the assumption of a 2D regional impedance to hold, since the twist angle estimation for the reduced data set at $T_2$ is not notably compromised. Both observations, the failed shear angle estimation for 1D data and the successful twist angle estimation for 3D data, lead to the conclusion that the impedance data of the DSM2 data set contains very strong noise that dominates the \emph{strike} single frequency, single site analysis. 

Assessing the same original data set, we performed the distortion analysis with \emph{gadget}. Observing Figures~\ref{fig:DSM2}g, \ref{fig:DSM2}h and \ref{fig:DSM2}i, \emph{gadget} retrieves almost all distortion angles with precision and accuracy, only the anisotropy angle $|\phi_a|>25^{\circ}$ exhibits increasing bias towards large angles (cp.~Figure \ref{fig:DSM2}i). Even though it appears that twist and shear angle estimations failed occasionally (cp.~labelled data in Figures~\ref{fig:DSM2}g to \ref{fig:DSM2}i), this failure can be linked to large original anisotropic distortion as we will explain in the following. The three angles are estimated simultaneously and form a triplet of parameters which represent the corresponding distortion, therefore the analysis result of each angle is linked to the others, which means that if one of them is estimated poorly, the estimation of the other two is likely to be poor as well. This happens because the computation of distortion optimality is based on the performance of the angle triplet and not on each angle separately. Figures~\ref{fig:DSM2}g, \ref{fig:DSM2}h and \ref{fig:DSM2}i clearly show that twist and shear angle are estimated poorly exactly when the original anisotropic distortion is large, whereas large original twist and shear angles are retrieved accurately when the anisotropic distortion is moderate. In this example, a critical value for original anisotropic distortion, above which the analysis results are compromised, is $|\phi_a|>30^{\circ}$. This critical value represents a gain factor of $\approx 3$ between the lower and the higher impedance off-diagonal components and it is likely based on the presence of noise as we will explain in the following. Anisotropic distortion is a site gain that increases or decreases the local horizontal electric field components in perpendicular directions and therewith, increases or decreases the respective impedance components. The impedance is formulated as a second rank tensor with, for real data, confidence limit estimates for each component. These confidence limits, even though for one specific component, are usually related to the larger impedance component (as it has been assumed in this specific, synthetic data set), which means that a particularly small impedance component is likely to have large confidence limits (representing low confidence in the estimate). When, i.e.~due to anisotropic distortion, the ratio between the smallest and largest components increases, the confidence in the estimate of the small components can drop below measurable. Such a situation heavily affects the impedance analysis which requires reasonable confidence in all estimates. In this example, $5\%$ of the maximum impedance component is the standard deviation of the added gaussian noise \citep{Miensopust:2013}, which, only considering anisotropic distortion $|\phi_a|=30^{\circ}$, increases three-fold as relative noise for the smaller components. Therefore, we interpret the critical value extracted from this example as a measure of the noise level in the present data and not as a general limit of the approach. From this test, we conclude that \emph{gadget} exhibits fair noise resistance and delivers reliable estimates in the presence of very strong noise and 1D data.

To illustrate the capability to treat 3D data, we performed the \emph{gadget} analysis on the reduced data set with $T_{\mathrm{min}}=10\,\mathrm{s}$, as discussed before for the \emph{strike} analysis. In Figures~\ref{fig:DSM2}j, \ref{fig:DSM2}k and \ref{fig:DSM2}l, we observe that the results obtained by \emph{gadget} are comparable to the results obtained from the analysis of the entire data set, but the confidence in the estimates has decreased considerably. The confidence in the estimates is, however, similar between sites that are far away from 3D features (the centre part of the model) and those that are close, thus, it is not generally possible to deduce that the penetration into 3D features is the cause of the confidence decrease. In fact, the similarity of confidence between the results of all sites suggests that 3D data can indeed be treated correctly, but that the large noise in the data affected it at periods larger than $10\,\mathrm{s}$ more strongly. This interpretation is also supported by our previous assessment that the critical angle up to which we can recover anisotropic distortion is descriptive for the present noise, because in this data the critical angle is somewhat lower, about $\approx 20^{\circ}$ (cp.~Figure~\ref{fig:DSM2}l).

\subsection{Real Data Set: BC87 (sites \emph{lit007}, \emph{lit008}, \emph{lit901} and \emph{lit902})}
The BC87 MT data set contains twenty-seven sites along a 150 km profile across resistive terrains in southeastern British Columbia \citep{Jones:1988,Jones:1993a,Ledo:2001}. These data are strongly distorted and have been analysed regularly to demonstrate distortion effects and analysis procedures \citep{Jones:1993b,Lilley:1993,Eisel:1993,deGroot:1995,Chave:1997,Ritter:1997,Gomez:2013a, Gomez:2013b}. Let us concentrate on four examples \emph{lit007}, \emph{lit008}, \emph{lit901} and \emph{lit902}, where the first two and the latter two form pairs of adjacent sites. \cite{Jones:1993b} report that, even though, the first two sites are as close as $2\,\mathrm{km}$, they exhibit fundamentally different impedances, which can only be explained by significant galvanic distortion. The authors note that, in contrast, the latter two are as similar as their proximity suggests for all but one impedance component. A detailed analysis of sites \emph{lit007} and \emph{lit008} is given by \cite{jones:2012b} with traditional methods that assume 2D regional impedances with the conclusion that the data must be considered 3D.

\begin{figure}[t]
	\centering
		\includegraphics[width=.48\textwidth]{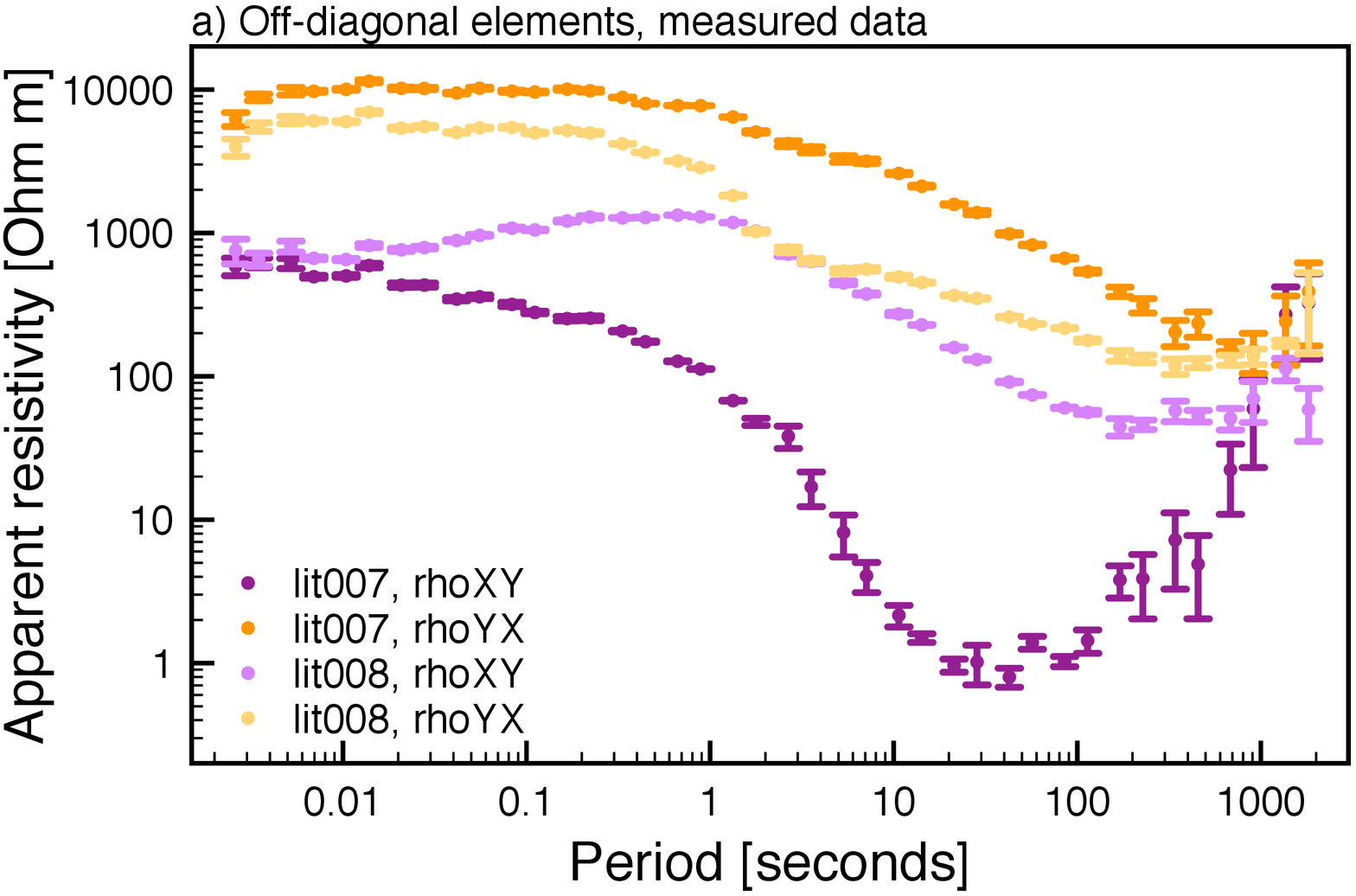}
		\includegraphics[width=.48\textwidth]{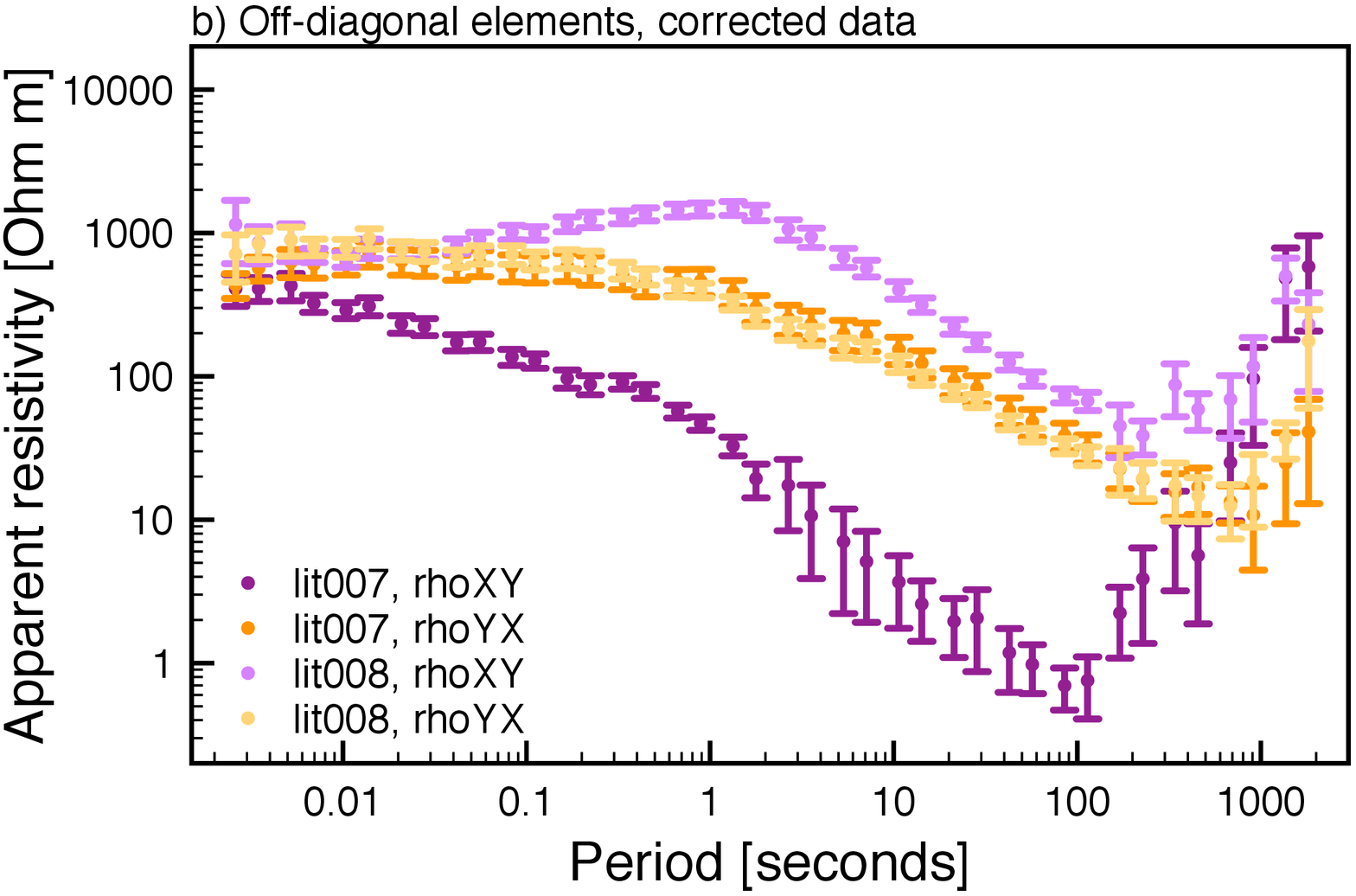}\\
		\includegraphics[width=.48\textwidth]{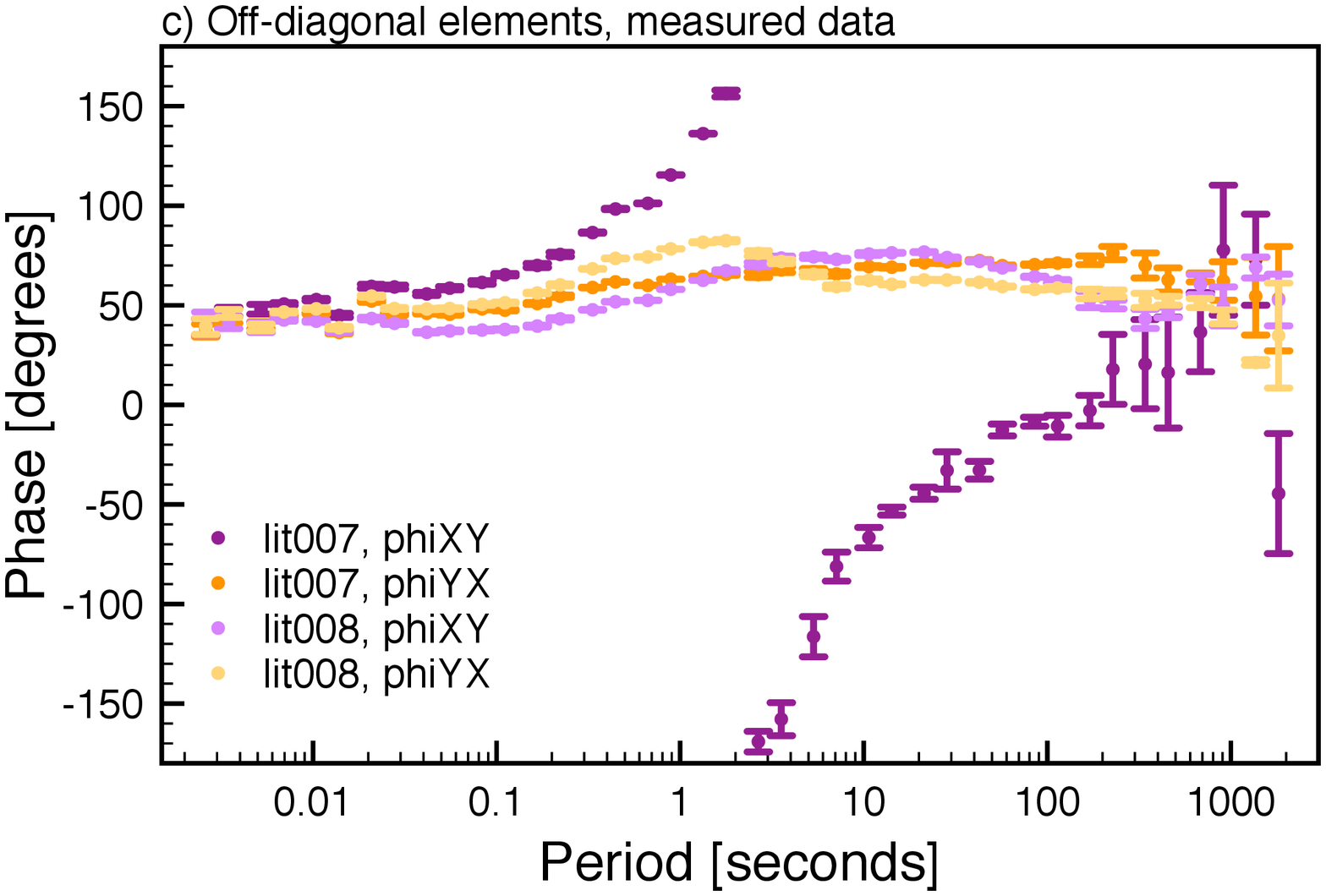}
		\includegraphics[width=.48\textwidth]{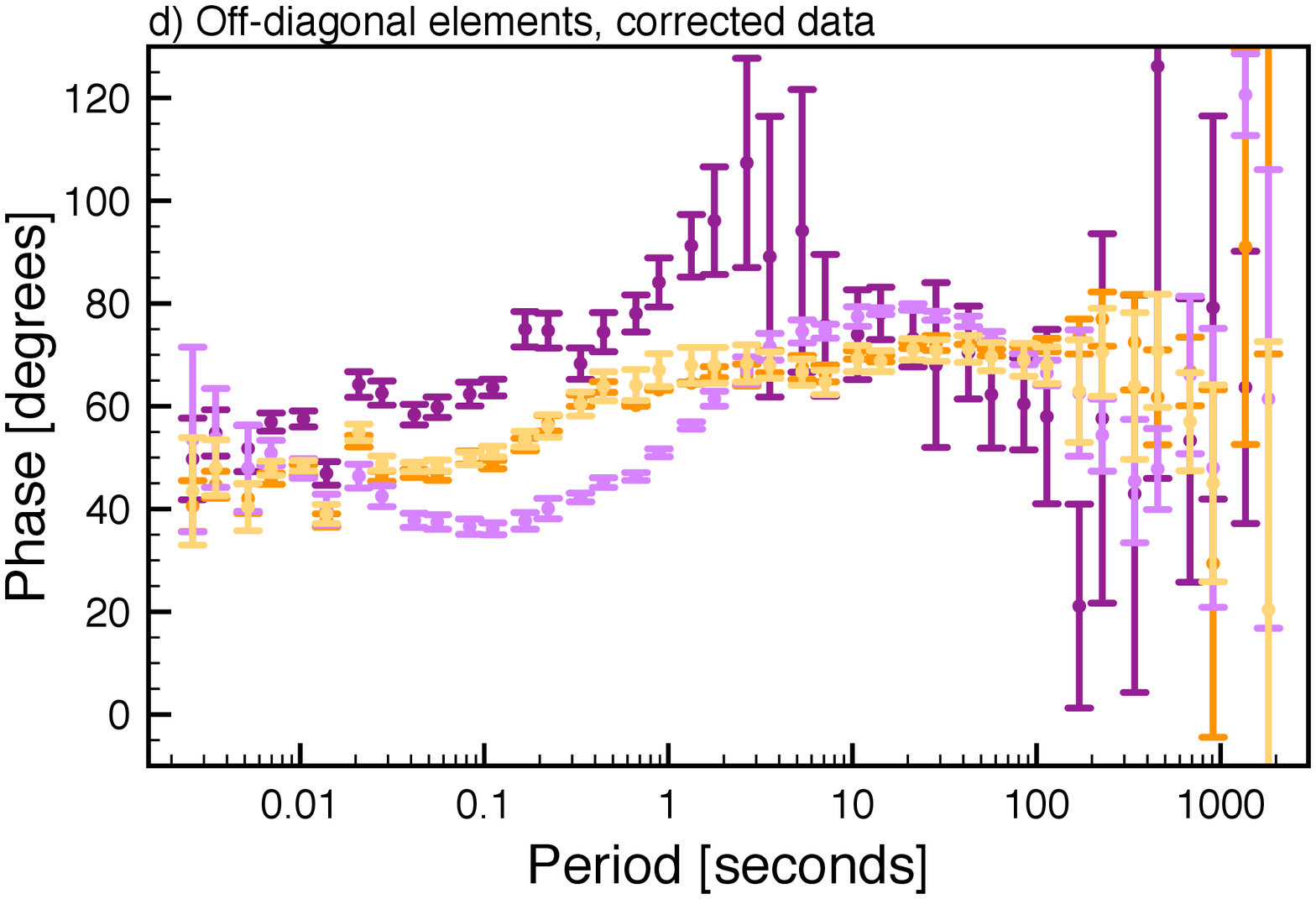}
	\caption{Apparent resistivity and phase data of sites \emph{lit007} and \emph{lit008} \citep{Jones:1993b} are illustrated. Despite their proximity (less than $2\,\mathrm{km}$ apart), the measured data of the two sites displays fundamentally different impedances, which hints at strong near-surface distortion effects. The distortion-corrected data reveals that the \emph{YX} components are virtually identically as would be expected for sites so close to each other. The \emph{TE} mode is more sensitive than the \emph{TM} mode to vertical resistivity contrasts, which could explain this observation \citep{swift:1971} and is supported by available geological information \citep{Jones:1993a}. }
	\label{fig:lit007008}
\end{figure}
Distortion analysis for sites \emph{lit007} and \emph{lit008} (see Figure \ref{fig:lit007008}) with \emph{gadget} reveals that the \emph{yx}-components are virtually identically as would be expected for sites that are so close to each other. Posteriori to the distortion analysis it was required to assume a galvanic shift factor of $g_{007}=3$ for site \emph{lit007}, in order to produce near identical apparent resistivity results for the \emph{Zyx}-component. This assumption is based on the observation that the \emph{a-posteriori} \emph{Zyx}-phases are identical between both sites within statistical confidence. The actual value of the static shift is, however, an arbitrary choice and only for an illustrative purpose to expose the virtual identity of the apparent resistivity components. 

\begin{table}[b]
\centering
\caption{Distortion analysis results of \emph{strike v5.0} and \emph{gadget v4.3} are compared. The distinct differences between both algorithms are caused by data three-dimensionality but nonetheless \emph{gadget} delivers geologically expected results for the pairs \emph{lit007}$/$\emph{lit008} and \emph{lit901}$/$\emph{lit902} as illustrated in Figures \ref{fig:lit007008} and \ref{fig:lit901902}. \newline}
\label{tab:lit}
\begin{tabular}{c | c || r@{$\,\pm\,$}l r@{$\,\pm\,$}l r@{$\,\pm\,$}l}
Site & Code & \multicolumn{2}{c}{Twist [degrees]} & \multicolumn{2}{c}{Shear [degrees]} & \multicolumn{2}{c}{Anisotropy [degrees]}\\  \hline\hline
\emph{lit007}& \emph{gadget}& 	$-13.19$&	$4.21$&	$-29.54$&	$5.15$&	$-23.78$&	$2.68$\\
\emph{lit007}& \emph{strike}& 		$10.47$&	$0.96$&	$25.04$&	$1.16$&	\multicolumn{2}{c}{NA}	\\ \hline
\emph{lit008}&\emph{gadget}&		$-15.70$&	$1.06$&	$-38.45$&	$1.89$&	$-16.30$&	$3.56$\\
\emph{lit008}&\emph{strike}&		$7.78$&	$1.58$&	$20.47$&	$2.33$&	\multicolumn{2}{c}{NA}	\\ \hline
\emph{lit901}&\emph{gadget}&		$-1.90$&	$0.63$&	$-6.20$&	$1.84$&	$0.21$&	$0.68$\\
\emph{lit901}&\emph{strike}&		$-3.33$&	$0.18$&	$-2.28$&	$0.35$&	\multicolumn{2}{c}{NA}	\\ \hline
\emph{lit902}&\emph{gadget}&		$6.36$&	$1.09$&	$4.97$&	$1.46$&	$-8.72$&	$1.81$\\
\emph{lit902}&\emph{strike}&		$4.04$&	$0.17$&	$6.06$&	$0.69$&	\multicolumn{2}{c}{NA}	
\end{tabular}
\end{table}
Since the \emph{TE} mode is much more sensitive than the \emph{TM} mode to the inductive effect of a vertical resistivity contrast \citep{swift:1971}, we interpret the finding of the identical $Z_{yx}$ components between the sites \emph{lit007} and \emph{lit008} as an indication of the presence of a vertical resistivity contrast in between them. This interpretation is supported by other studies \citep{Jones:1988,Jones:1993a} that recognise an interface of two geological units between the two sites, but, to date, this interpretation could not been deduced convincingly from the measured MT impedance data due to the limitation in distortion analysis to 2D regional impedance data. The distortion analysis results are summarised in Table \ref{tab:lit} and compared in between the algorithms \emph{strike v5.0} and \emph{gadget v4.3}. The large differences between results of the two algorithms indicate the presence of a highly 3D subsurface at both sites with strong electric galvanic distortion. \cite{Chave:1997} report  similar shear distortion parameters (at the shortest periods) to our results and larger twist angles, but the authors note that the data cannot fit the 2D impedance model required by their analysis method and thus, they conclude that it must be considered 3D. The study by \cite{Chave:1997} also includes the analysis for magnetic field distortion, with which the data still could not be fitted to a 2D model, and further strengthens the suggestion of present 3D induction effects. 

\begin{figure}[t]
	\centering
		\includegraphics[width=.48\textwidth]{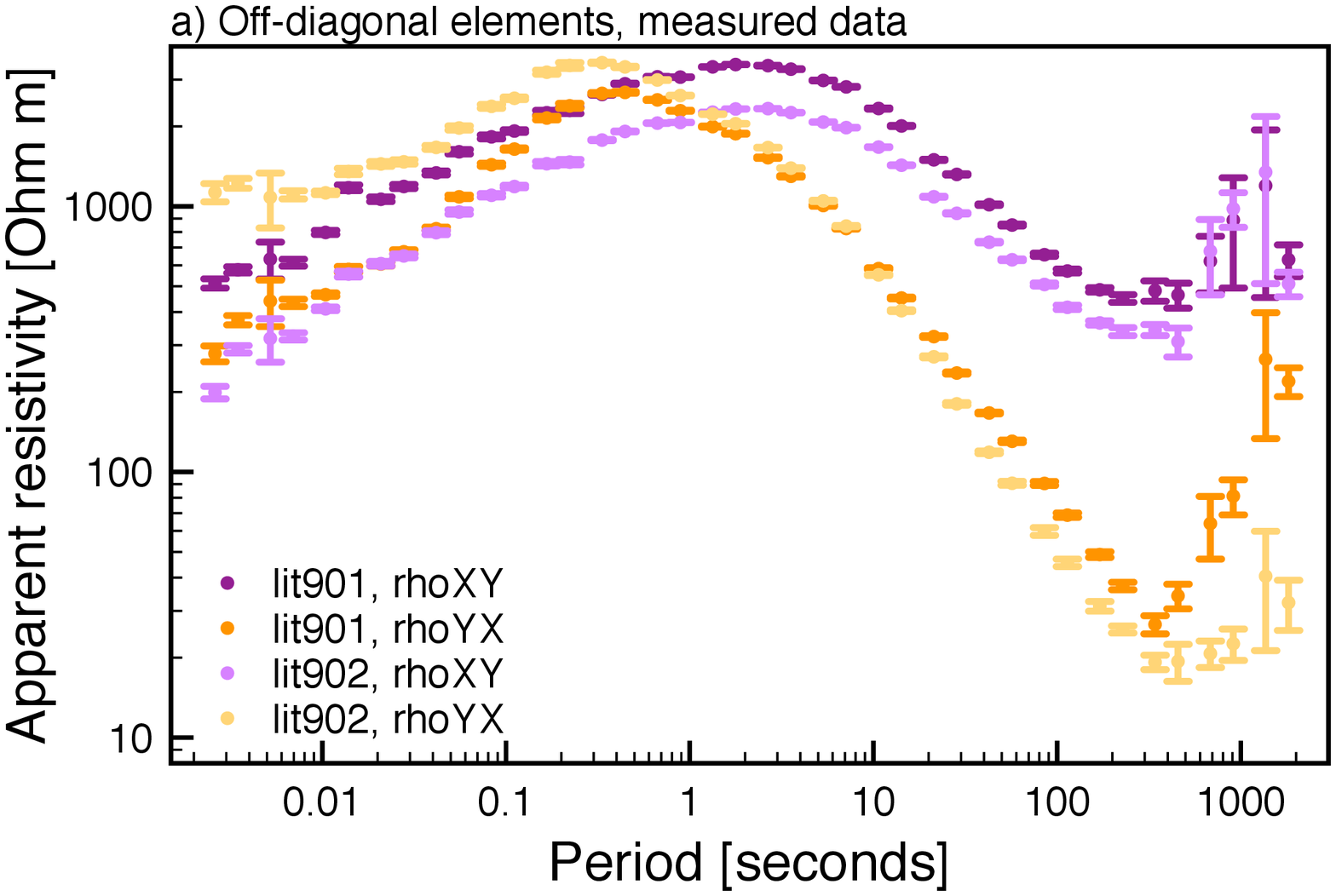}
		\includegraphics[width=.48\textwidth]{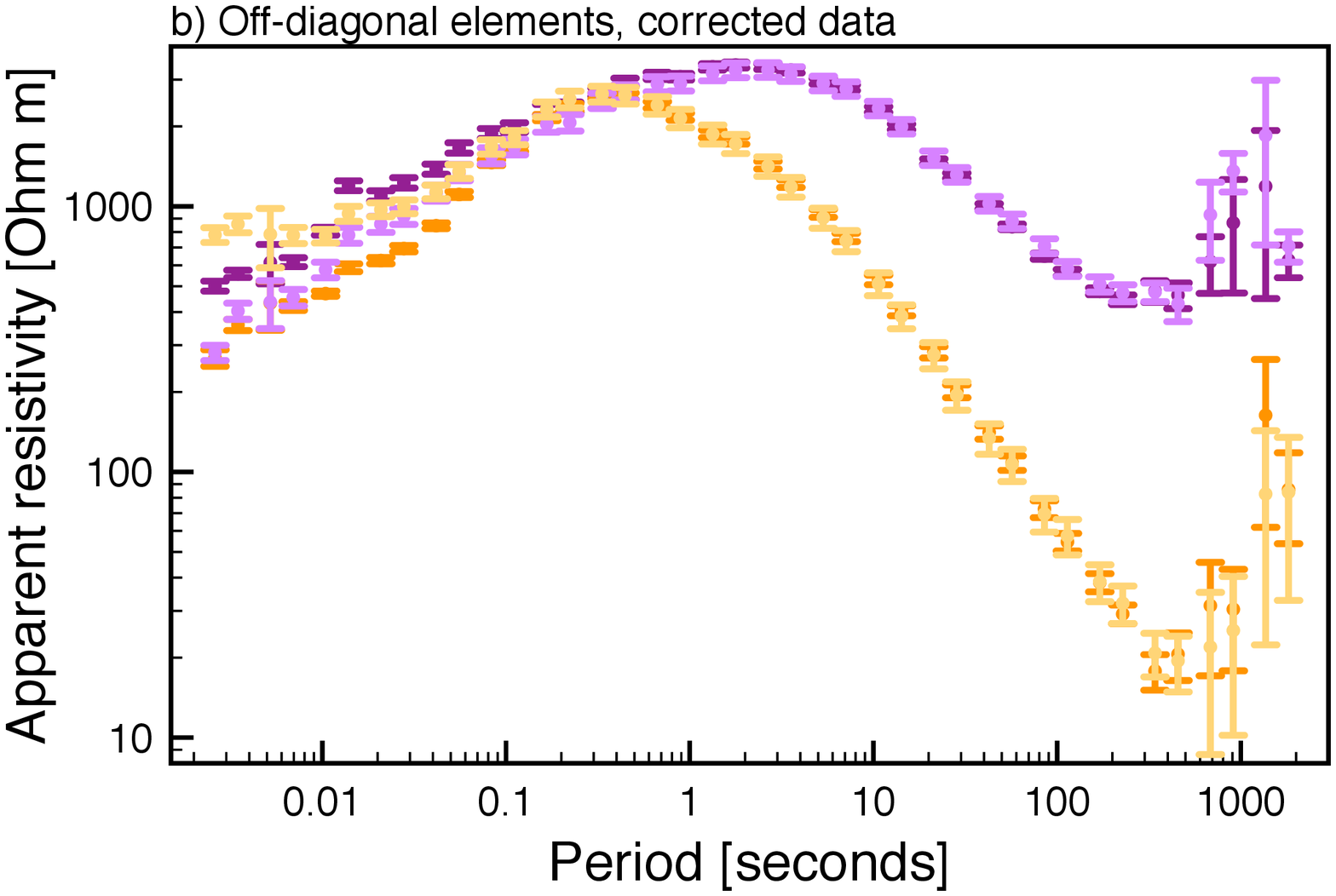}\\
		\includegraphics[width=.48\textwidth]{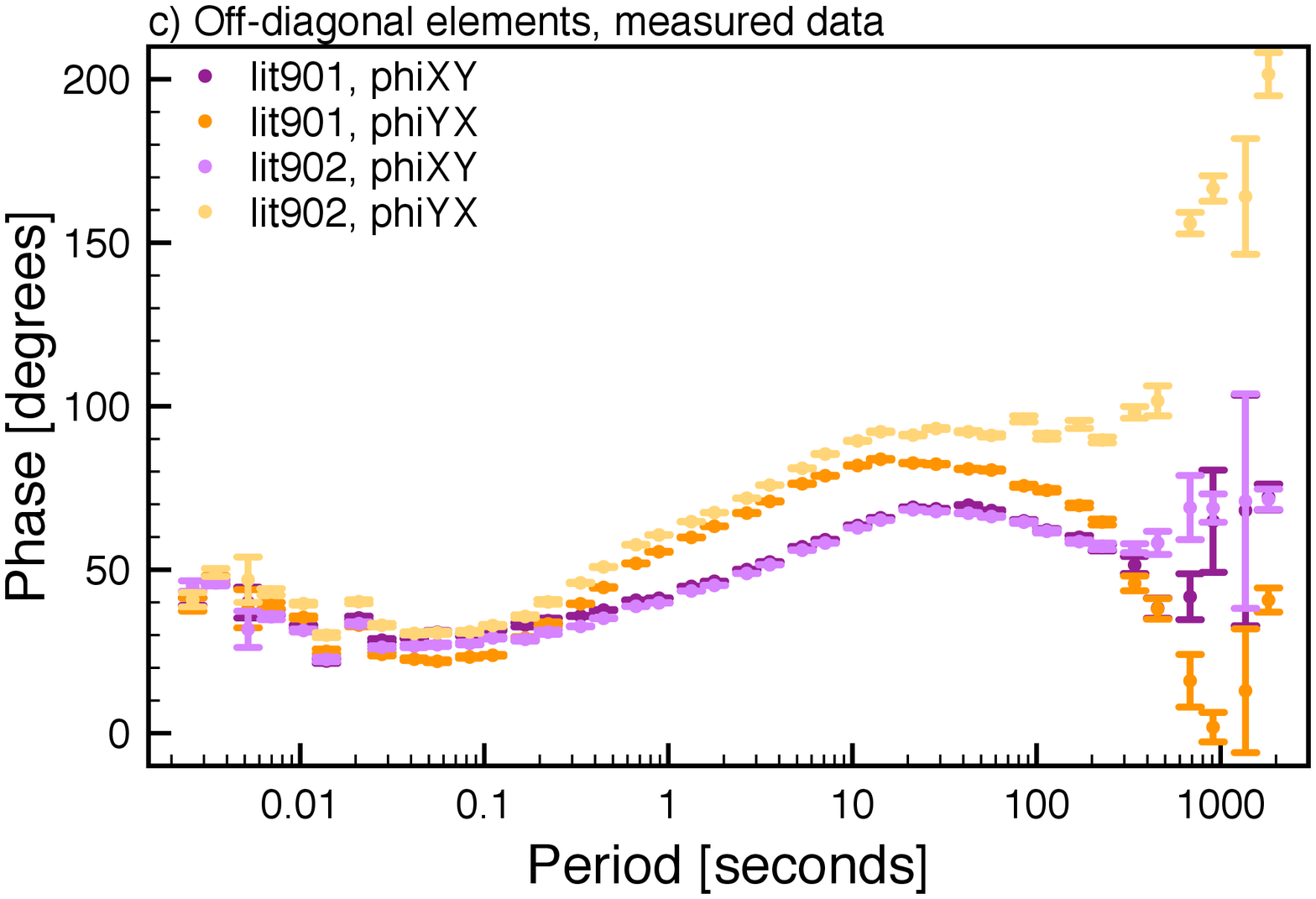}
		\includegraphics[width=.48\textwidth]{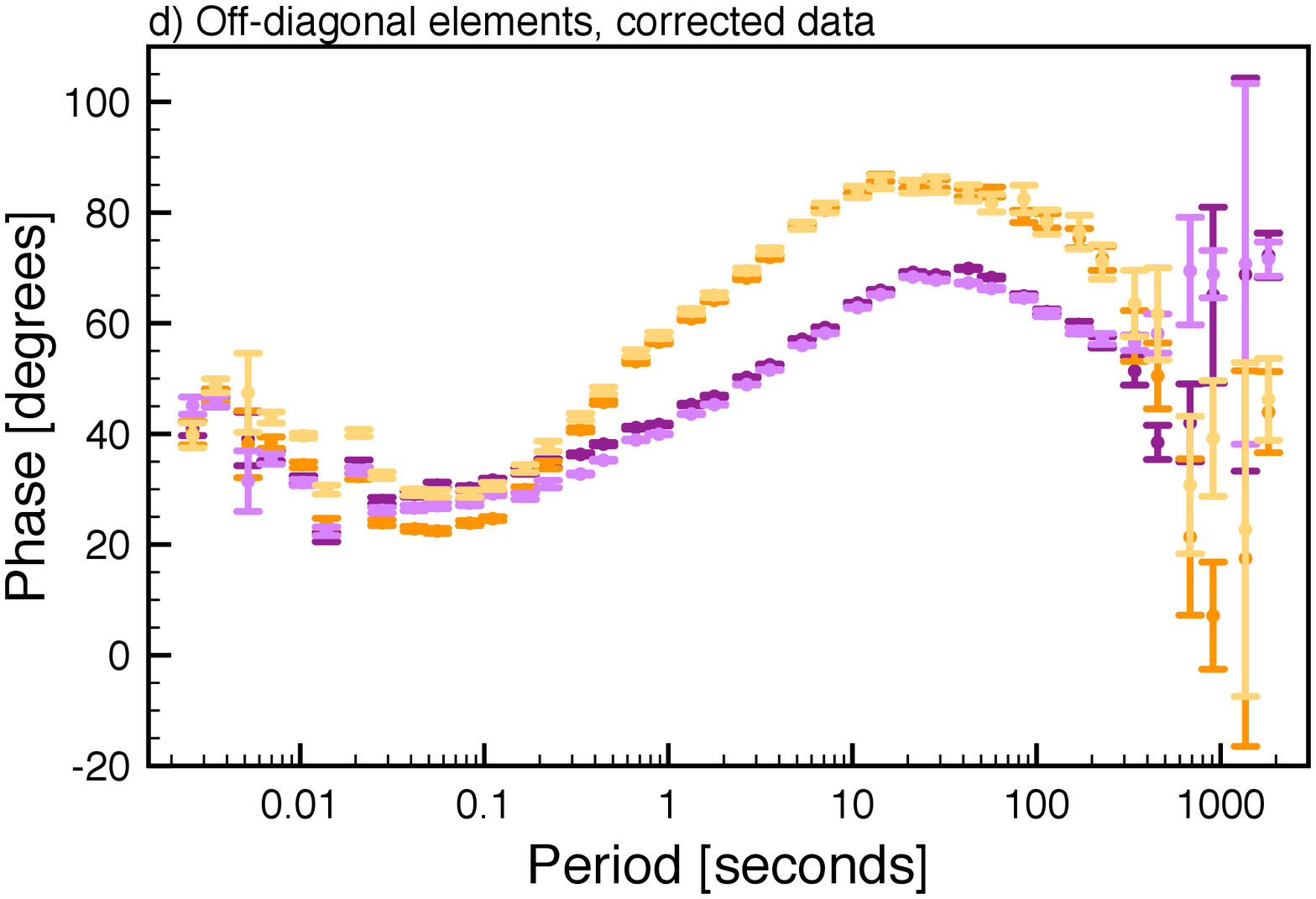}
	\caption{Apparent resistivity and phase data of sites \emph{lit901} and \emph{lit902} \citep{Jones:1993b} are illustrated. Despite their proximity, the measured data of the two sites displays slightly different impedances indicating near-surface distortion effects. The distortion-corrected data reveals that all components are near identical as expected since both sites are located on the same geological formation \citep{Jones:1993a,Jones:1993b}.}
	\label{fig:lit901902}
\end{figure}
Sites \emph{lit901} and \emph{lit902}, analysed in Figure \ref{fig:lit901902}, yield near identical impedance off-diagonal elements after distortion correction and confirm the proximity and geological similarity between the sites. As for the previous example, a galvanic shift correction of the apparent resistivity curves is required to match the observation in the phase (unaffected by galvanic shift) and the corrected resistivity data. The empirical gain factors are $g_{901\mathrm{, xy}}=1$, $g_{901\mathrm{, yx}}=1.5$, $g_{902\mathrm{, xy}}=2.75$ and $g_{902\mathrm{, yx}}=1$. The distortion analysis results are summarised in Table \ref{tab:lit} and compared in between the algorithms \emph{strike v5.0} and \emph{gadget v4.3}. The moderate differences between the algorithms indicate the presence of a 3D subsurface at both sites with small electric galvanic distortion. \cite{Jones:1993b} find that, once a 3D/2D regional model for the impedance of site \emph{lit902} is adopted, frequency dependent twist and shear angles can be recovered within statistical confidence, which are similar (but not identical) to ours at the shortest periods (see \emph{strike} results in Table \ref{tab:lit}). The authors don't arrive to the same conclusion for site \emph{lit901} and remain sceptical towards the observed differences between the \emph{yx}-components of the two adjacent sites. \cite{Chave:1997} report similar shear distortion parameters (for both sites at the shortest periods) to our results but larger twist angles and acknowledge that the data of both sites cannot sufficiently fit a 2D regional model assuming frequency independent distortion and that thus, 3D inductive effects are likely present. These two examples demonstrate that our proposed analysis scheme waives the assumption of a 2D regional impedance and therefore recovers impedances that would be expected, given the available geological information. 

However, \emph{gadget} does not estimate sufficiently accurate anisotropic distortion and therefore it was necessary to empirically correct for galvanic shift \citep[cp.][]{Neukirch:2016a}. We argue, however, that the algorithm finds the larger part of the anisotropic distortion, as demonstrated in section \ref{sec:EM3DII}, which may help in assessing the data if there is no alternative available, e.g.~another nearby site or independent results from another method. Also note that the applied galvanic shift correction only serves for illustrative purposes, to demonstrate evident similarity between each pair of sites and we particularly disclaim correctness of the employed galvanic shift factors, because our proposed method cannot determine this parameter as we stated before.

\section{Conclusions}
We propose a new galvanic electric distortion analysis based on the MT Amplitude-Phase Tensor decomposition and we present our developed algorithm for this purpose. Our assumption is that, since the Phase Tensor is unaffected by galvanic distortion, all distortion must be present in the Amplitude Tensor. Both the Amplitude and Phase Tensors must contain the inductive, geometric information of the subsurface and thus, they are physically coupled and geometrically similar. Based on this coupling, we propose an optimisation strategy that maximises similarity between Amplitude and Phase Tensor geometric parameters and recovers the distortion tensor up to a
single constant, usually denoted as galvanic shift or site gain $g$.

On a synthetic data set, we illustrate the distortion effect on the Amplitude Tensor by manually applying distortion. These observed distortion effects are reverse engineered to define a weighted multi-objective
function that minimises when the corrected Amplitude Tensor, in comparison to the Phase Tensor, is distortion free. We propose to use a genetic algorithm, discuss the sub-objective functions and necessary weighting to find the optimal galvanic electric distortion parameters. Additionally, we demonstrate that the presented distortion analysis is independent of dimensionality
assumptions and thus, is suitable for 3D data. On a large scale synthetic data study (144 sites), we find that the proposed methodology delivers reliable results, even for data with a clear 3D
regional structure. 

Lastly, we test the performance of the proposed methodology on real data of two site pairs of the data set BC87, the adjacent sites \emph{lit007}/\emph{lit008} and \emph{lit901}/\emph{lit902}. 
We are able to recover impedances of this data set that are in accordance to geologic information. These results highlight the power of the presented distortion analysis since it is the first time, at the knowledge of the authors, that this has been achieved.
The explanation on why previous studies have not been successful in recovering interpretable impedances, is that the data contains significant 3D inductive effects that cannot be analysed by state of the art methodologies; a limitation that is overcome by our algorithm.


\section{Supplements}
The algorithm described in this work is available as a MatLab executable under:\begin{center}\url{https://www.dropbox.com/sh/p34oxege5e8qmoa/AACldERjVFUWoPRR4cKips_4a?dl=0}\end{center}
for academic users.

\section{Acknowledgments}
This work was partly funded by Repsol under the framework of the CO-DOS project. We want to thank Marion Miensopust for providing the distortion parameters for the 3D dataset from the MT3D-2 workshop. All MT3D workshop datasets are publicly available at \begin{center} \url{http://www.complete-mt-solutions.com/mtnet/workshops/em_workshops.html#3DMTINV} \end{center} for download. The constructive criticism of the reviewers John Booker, Alan Jones, Anna Avdeeva and an anonymous reviewer, and the editors Gary Egbert and Ute Weckmann have greatly improved earlier versions of this paper. Our special thanks are addressed to Anna Avdeeva, whose suggestions inspired the data adaptive variance weighting of the sub-objective functions.

\appendix

\section{Galvanic Electric Distortion and the Amplitude Tensor}
\label{sec:GalvanicDistortion}
%

Galvanic distortion of the electric field can be described by $\mathbf{E}_d = \mathbf{C}   \mathbf{E}$
\citep[][and references therein]{jones:2012b} and inserting distortion into $\mathbf{E}=\mathbf{Z}   \mathbf{H}$ yields the relation that is observed by field measurements:
 \begin{equation}
\mathbf{E}_d = \mathbf{C}   \mathbf{E} = \mathbf{C}   \mathbf{Z}   \mathbf{H} = \mathbf{Z}_d   \mathbf{H},
\end{equation} 
with the observed MT impedance $\mathbf{Z}_d= \mathbf{C}   \mathbf{Z}$ related to the desired
(regional) MT impedance $\mathbf{Z}$. For the parameterisation of the real-valued
$2\times2$ distortion tensor $\mathbf{C}$, let us adopt the decomposition of \cite{Groom:1989}
and describe it with four unique parameters,
\begin{enumerate*}
\item[$\!$] a site gain $g\in\mathbb{R}^+$,
\item[$\!$] a twist angle $\phi_t\in[-\frac{\pi}{2},\frac{\pi}{2}]$,
\item[$\!$] a shear angle $\phi_s\in[-\frac{\pi}{4},\frac{\pi}{4}]$ and
\item[$\!$] an anisotropy parameter $a\in[-1,1]$,
\end{enumerate*}
where, without loss of generality, we express $a=\tan\phi_a, \phi_a\in[-\frac{\pi}{4},\frac{\pi}{4}]$, in angular form:
 \begin{equation}
\mathbf{C}=
  \begin{pmatrix}
  b & c\\
  d & e
  \end{pmatrix}
  = g' \mathbf{T}   \mathbf{S}   \mathbf{A}=
  g'  
  \begin{pmatrix}
  1 & \tan \phi_t\\
  -\tan \phi_t & 1
  \end{pmatrix}
  \begin{pmatrix}
  1 & \tan \phi_s\\
  \tan \phi_s & 1
  \end{pmatrix}
  \begin{pmatrix}
  1+\tan\phi_a & 0\\
  0 & 1-\tan\phi_a
  \end{pmatrix},
\label{eq:DistortionMatrix}
\end{equation} 
with $g'=\frac{g}{\sqrt{\det(\mathbf{T})\det(\mathbf{S})\det(\mathbf{A})}}$ including the normalisation factors for $\mathbf{T}$, $\mathbf{S}$ and $\mathbf{A}$ \citep{Groom:1989}. Throughout this manuscript, we assume $g'=1$, which means that the galvanic shift parameter is absorbed into the determinant of the impedance.


\subsection{Distortion Effects on Amplitude Tensor Parameters}
\begin{figure}
	\centering
		a) Twist (left), shear (middle) and anisotropic (right) distortion affecting tensor parameters.\\
		\includegraphics[width=.325\textwidth]{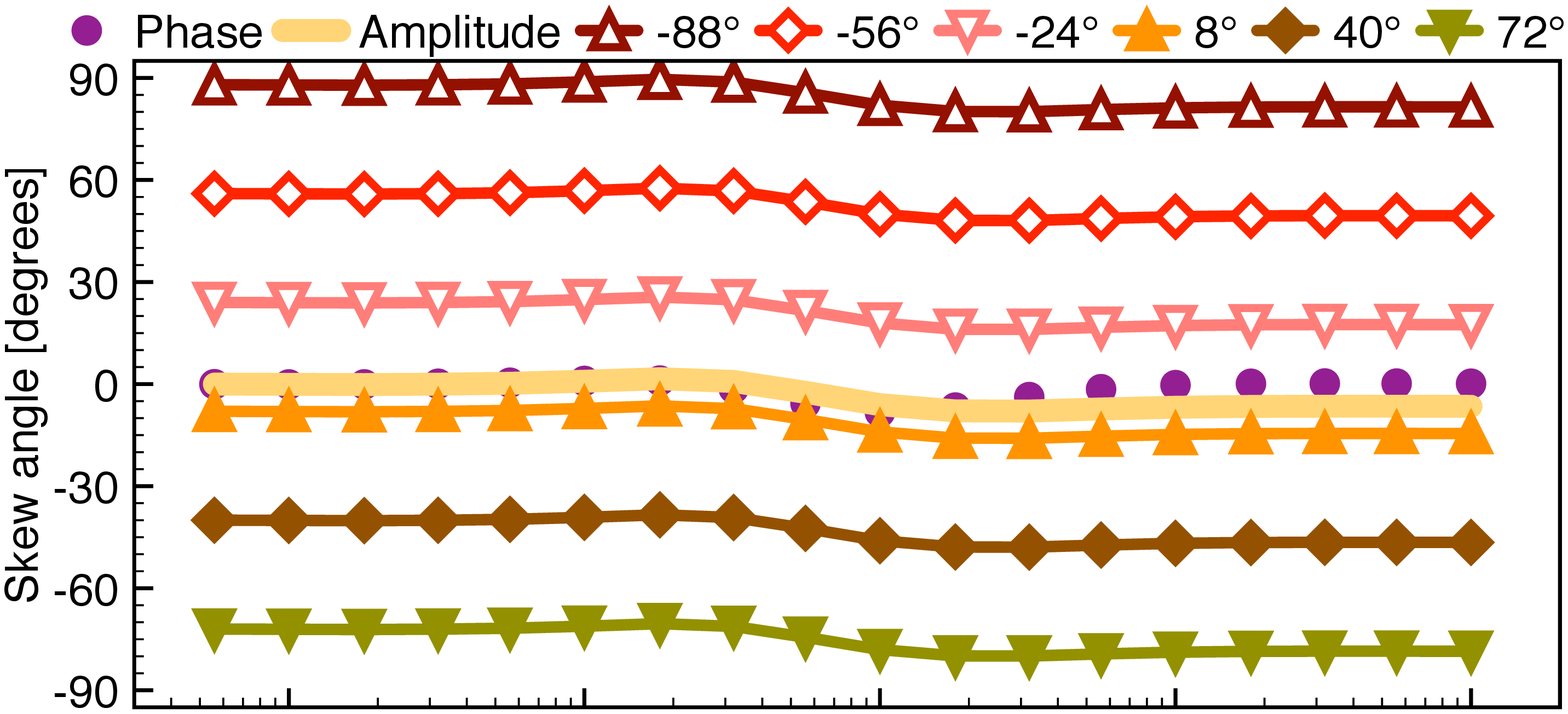}
		\includegraphics[width=.325\textwidth]{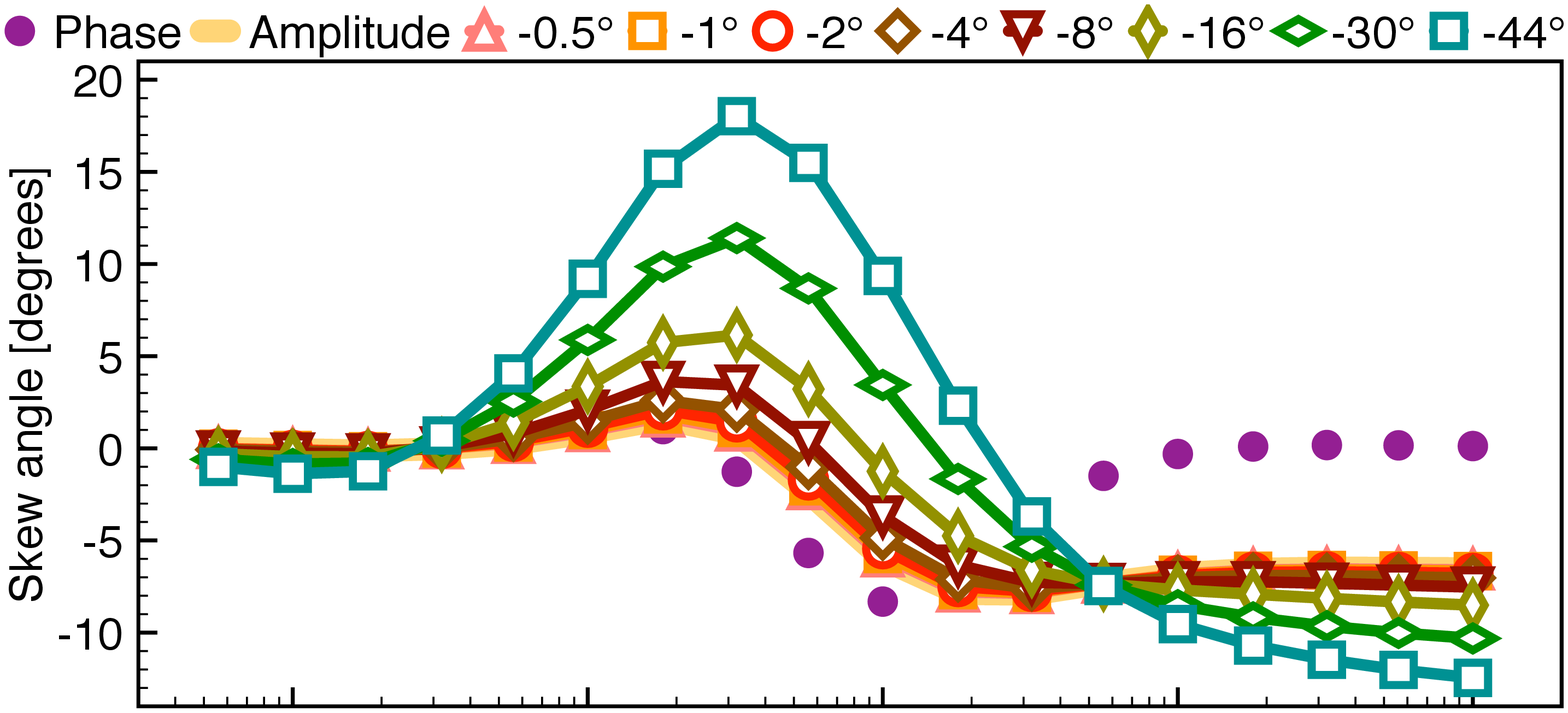}
		\includegraphics[width=.325\textwidth]{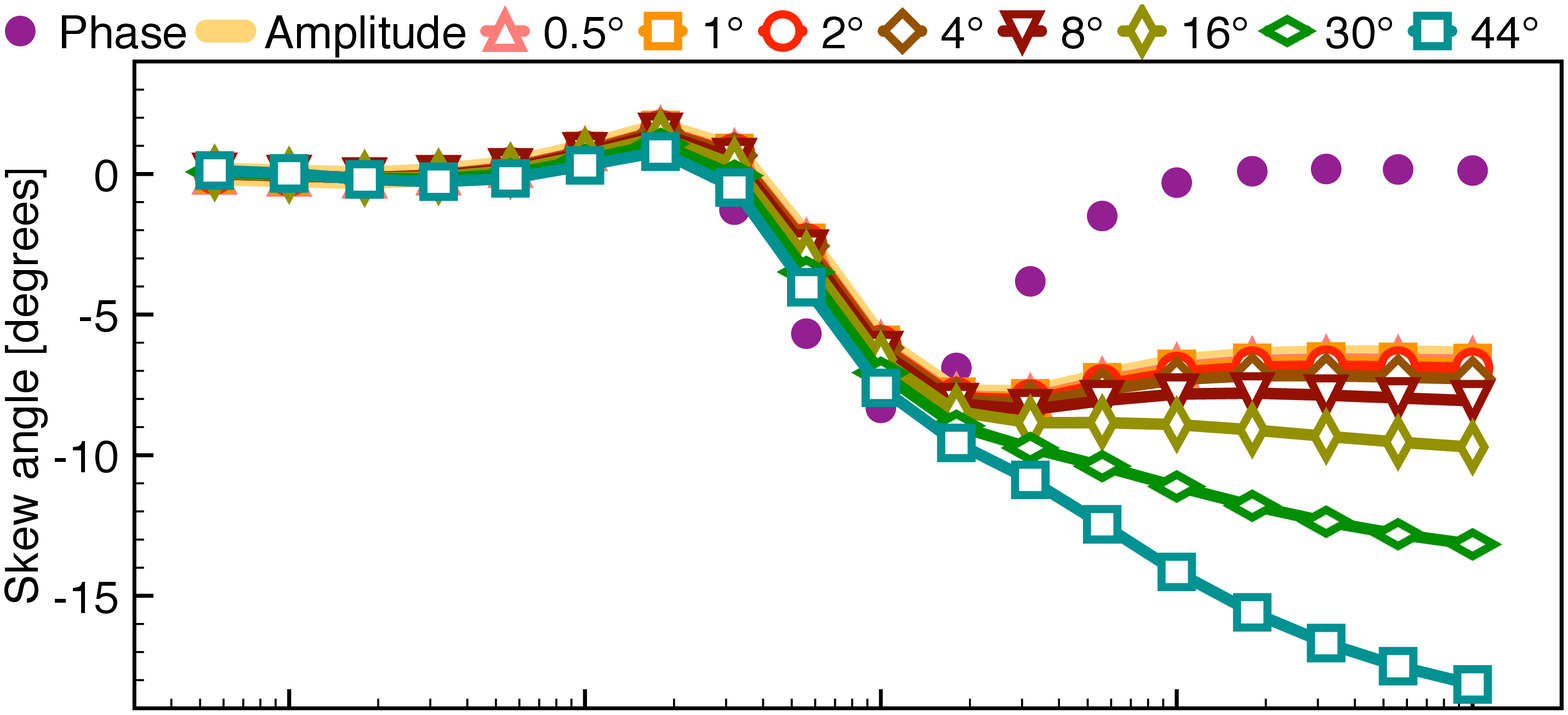}\\
		\includegraphics[width=.325\textwidth]{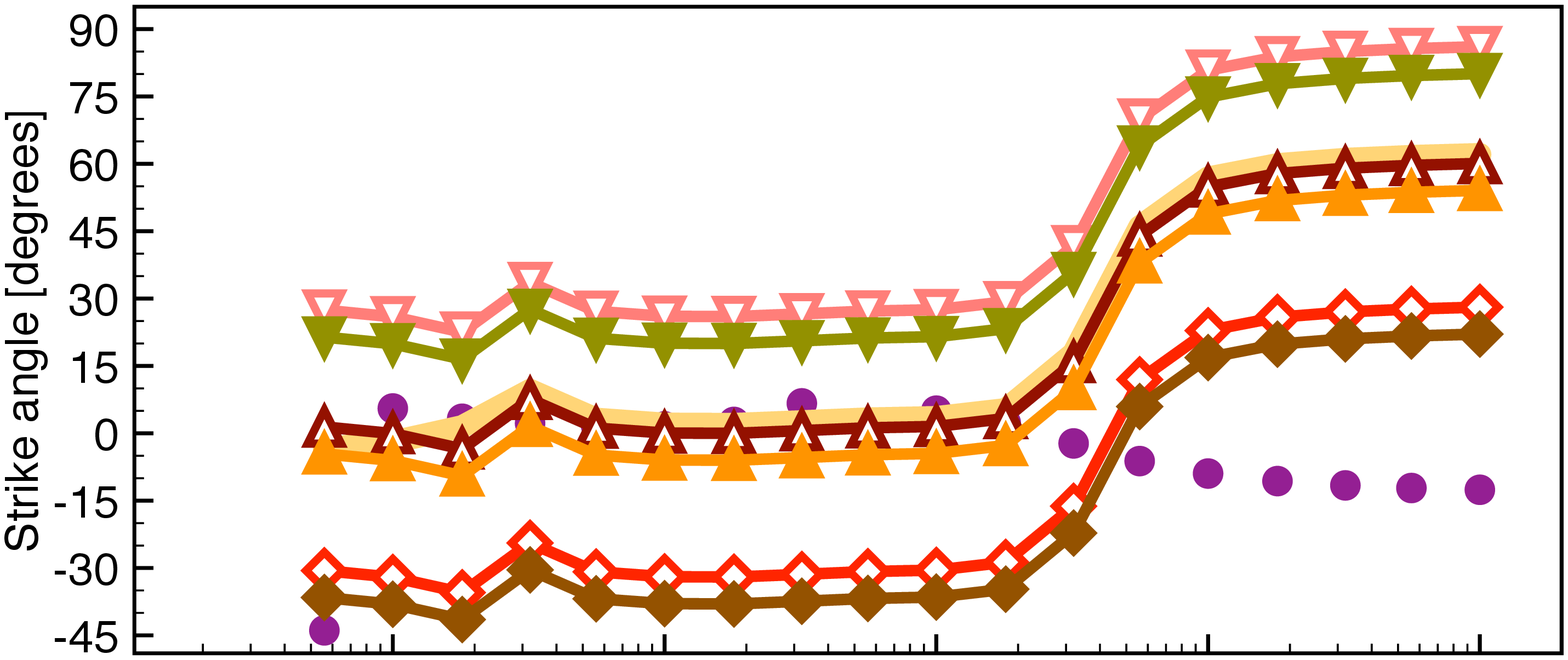}
		\includegraphics[width=.325\textwidth]{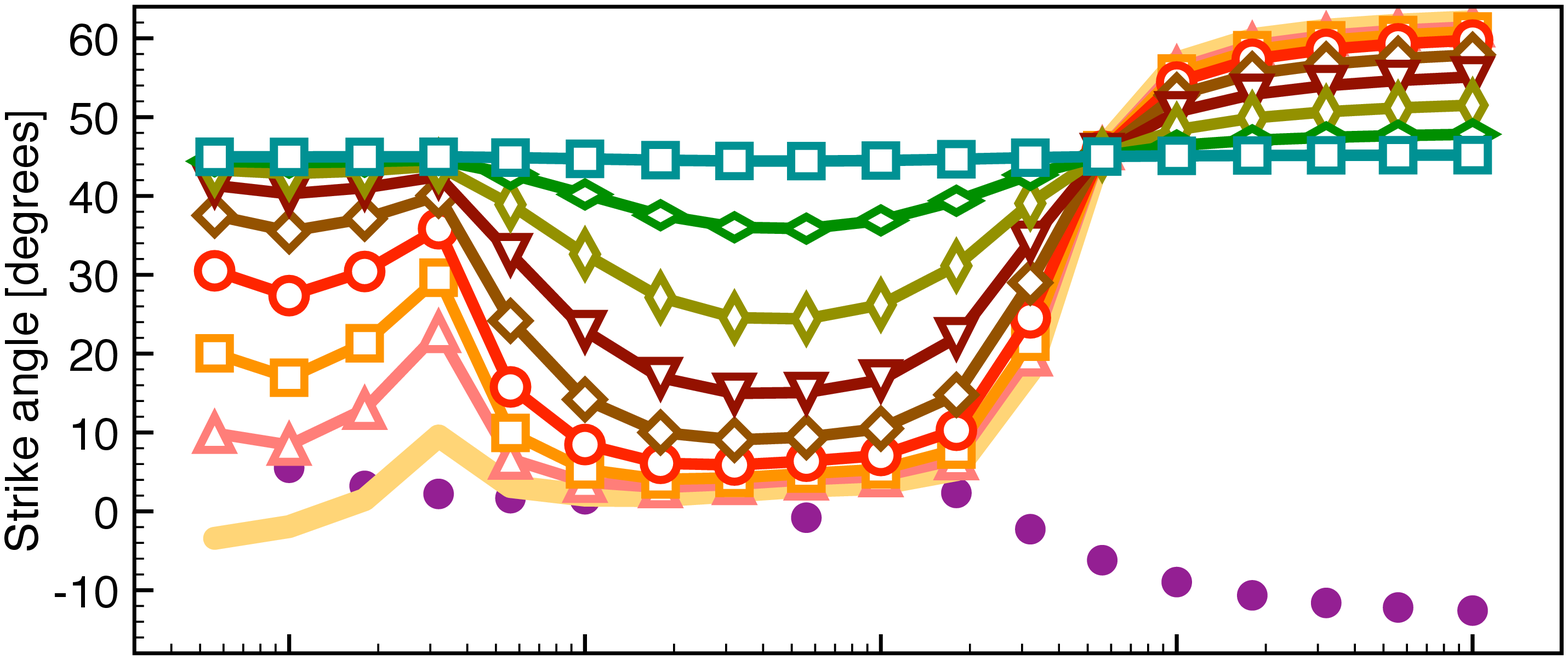}
		\includegraphics[width=.325\textwidth]{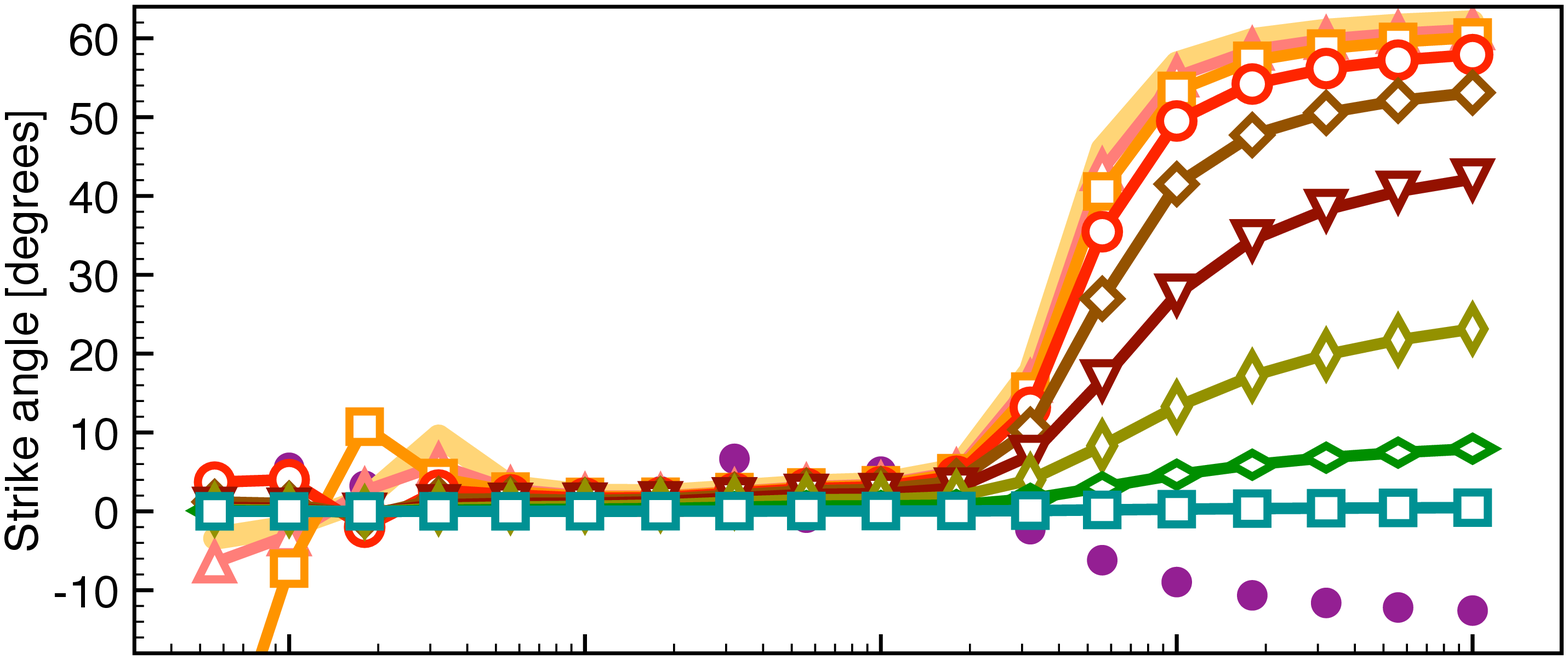}\\
		\includegraphics[width=.325\textwidth]{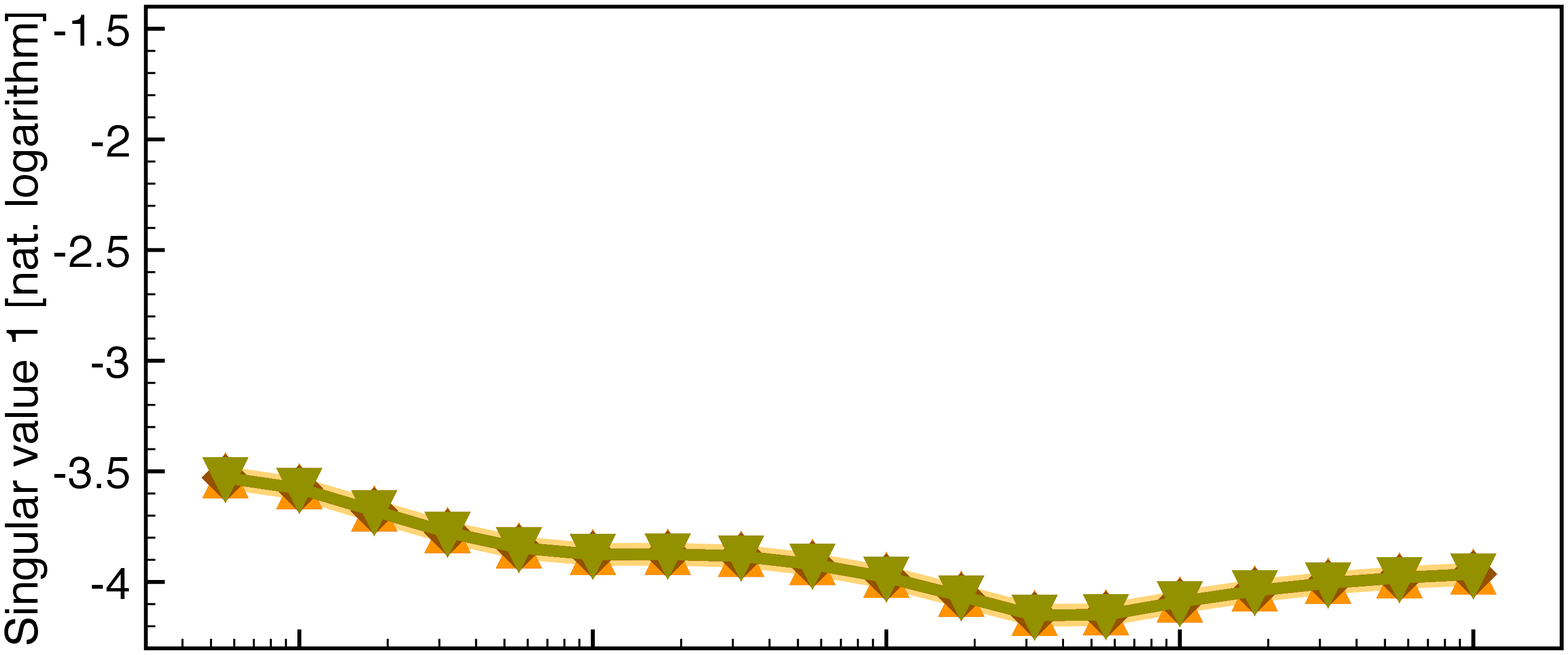}
		\includegraphics[width=.325\textwidth]{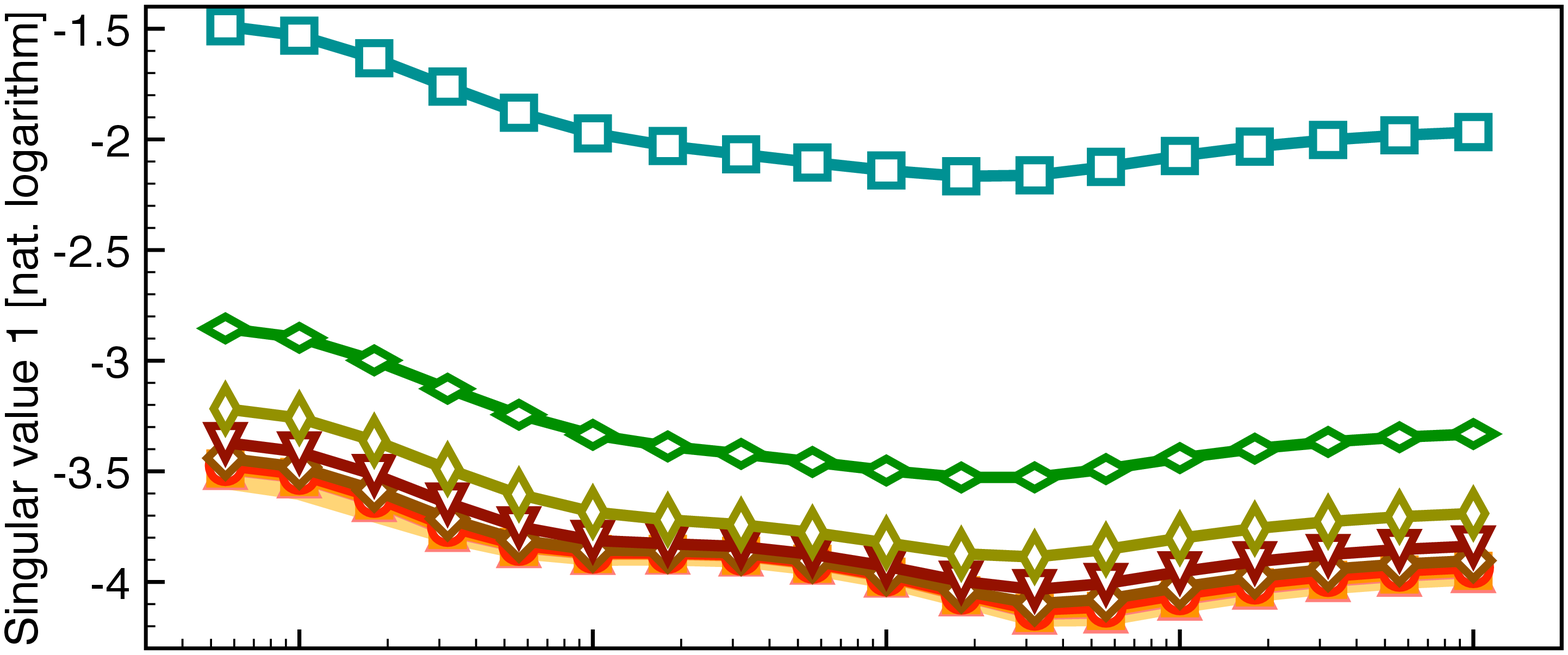}
		\includegraphics[width=.325\textwidth]{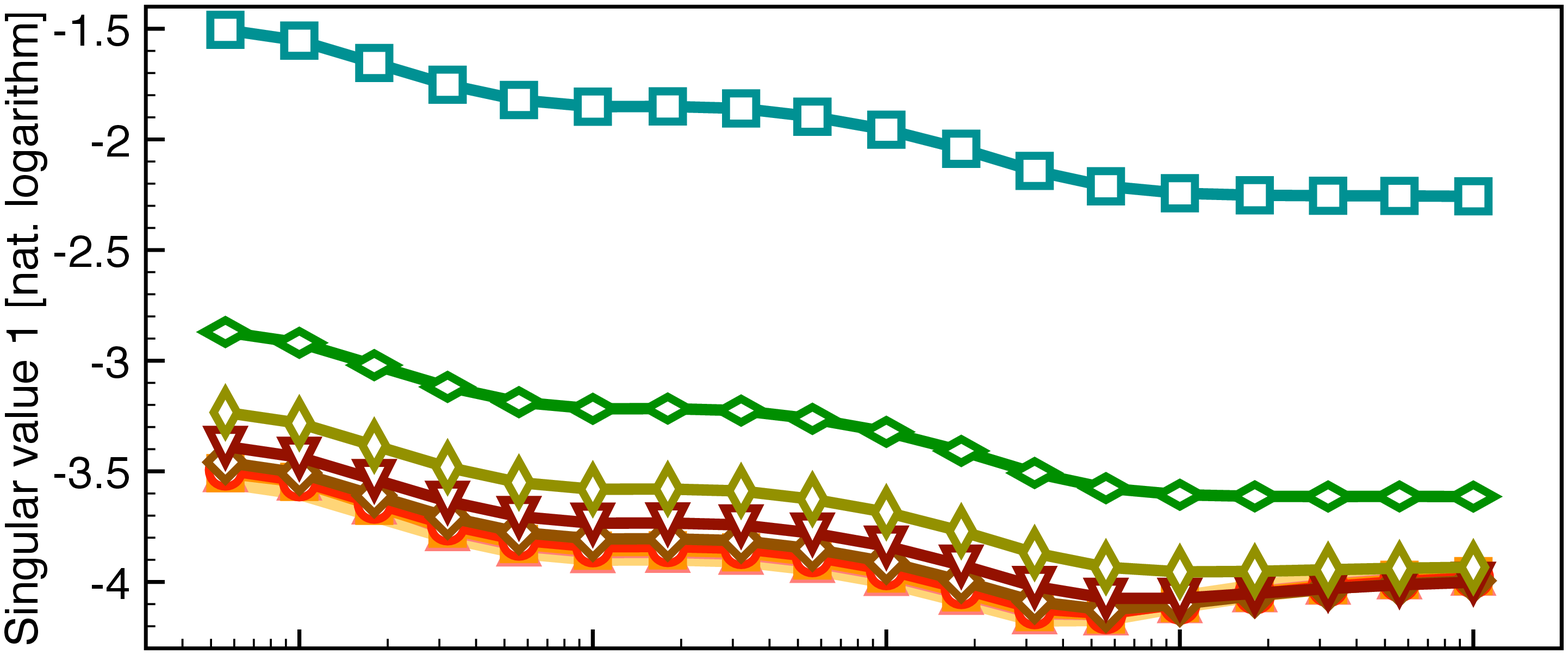}\\
		\includegraphics[width=.325\textwidth]{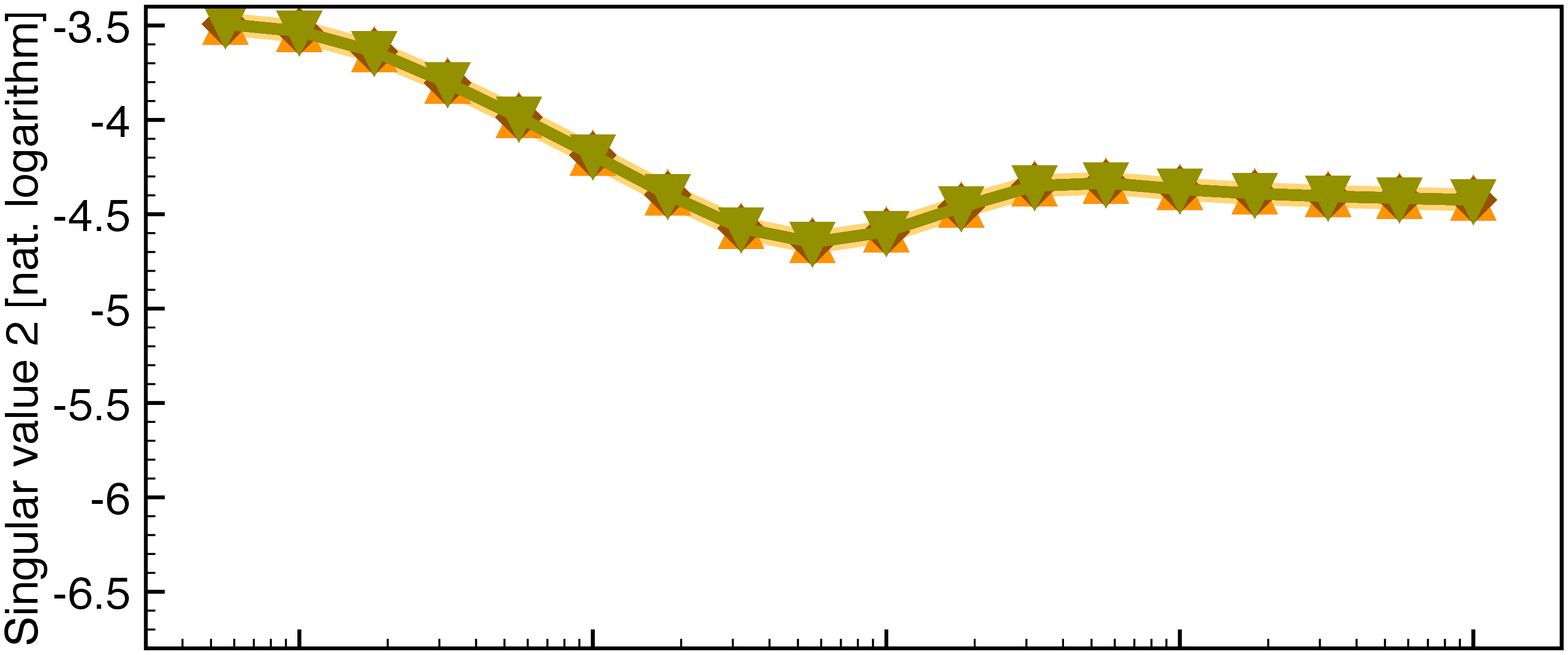}
		\includegraphics[width=.325\textwidth]{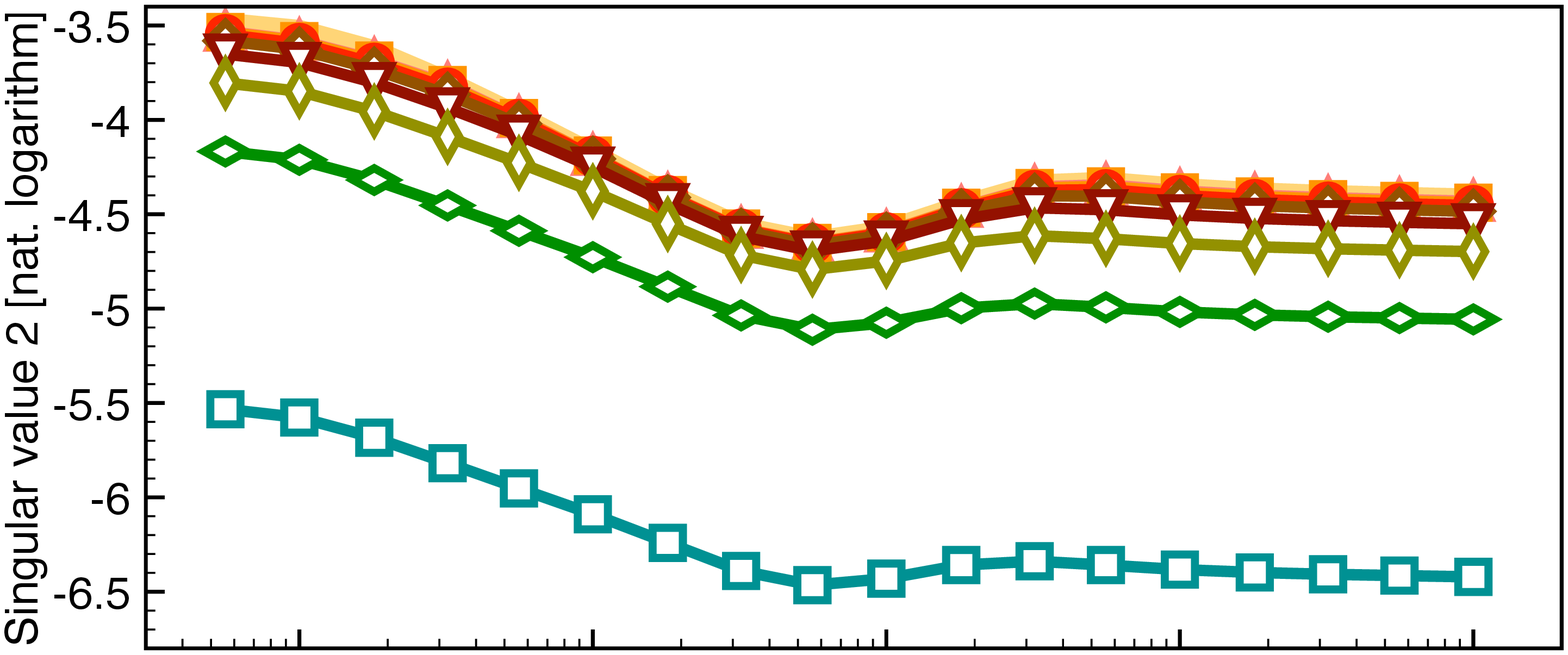}
		\includegraphics[width=.325\textwidth]{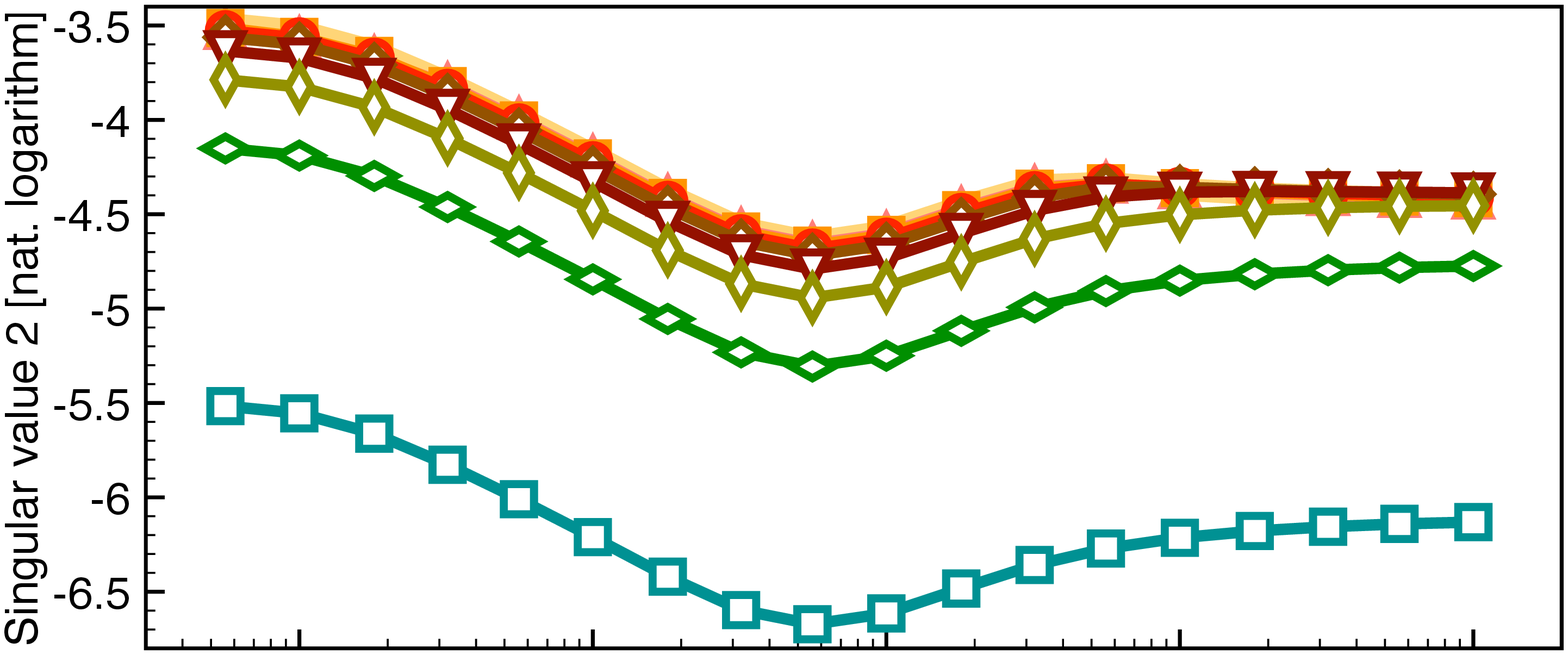}\\
		\includegraphics[width=.325\textwidth]{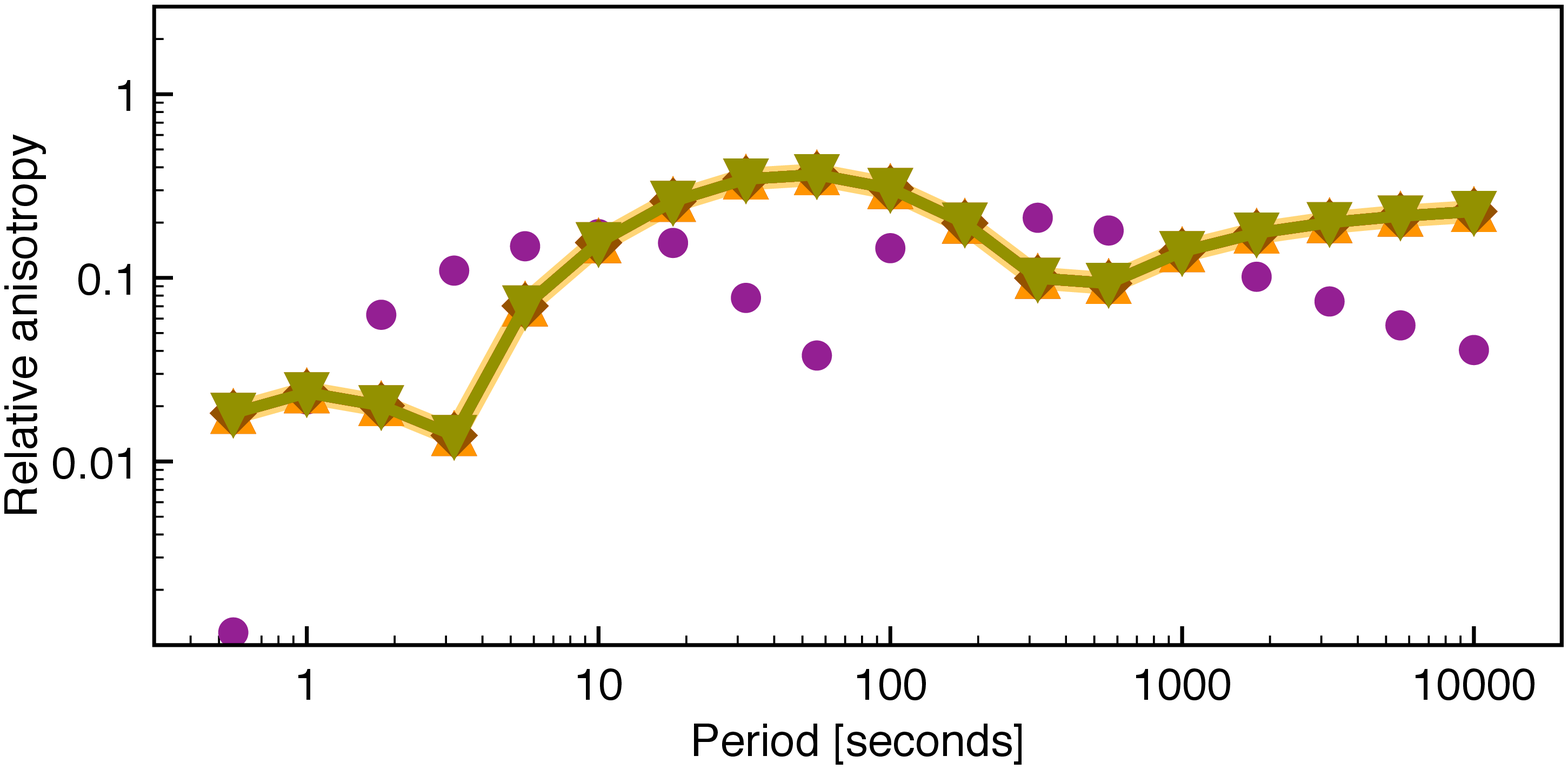}
		\includegraphics[width=.325\textwidth]{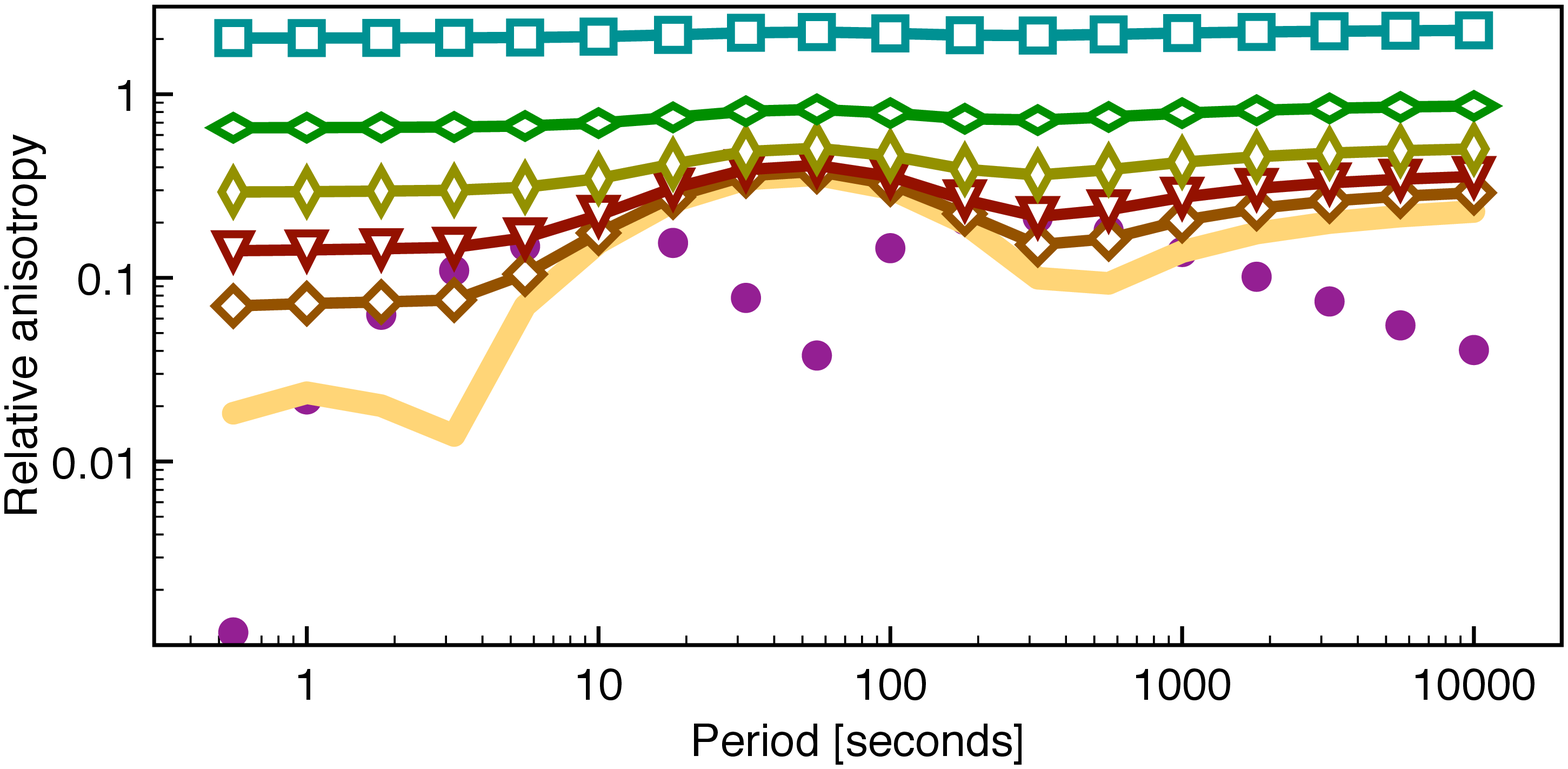}
		\includegraphics[width=.325\textwidth]{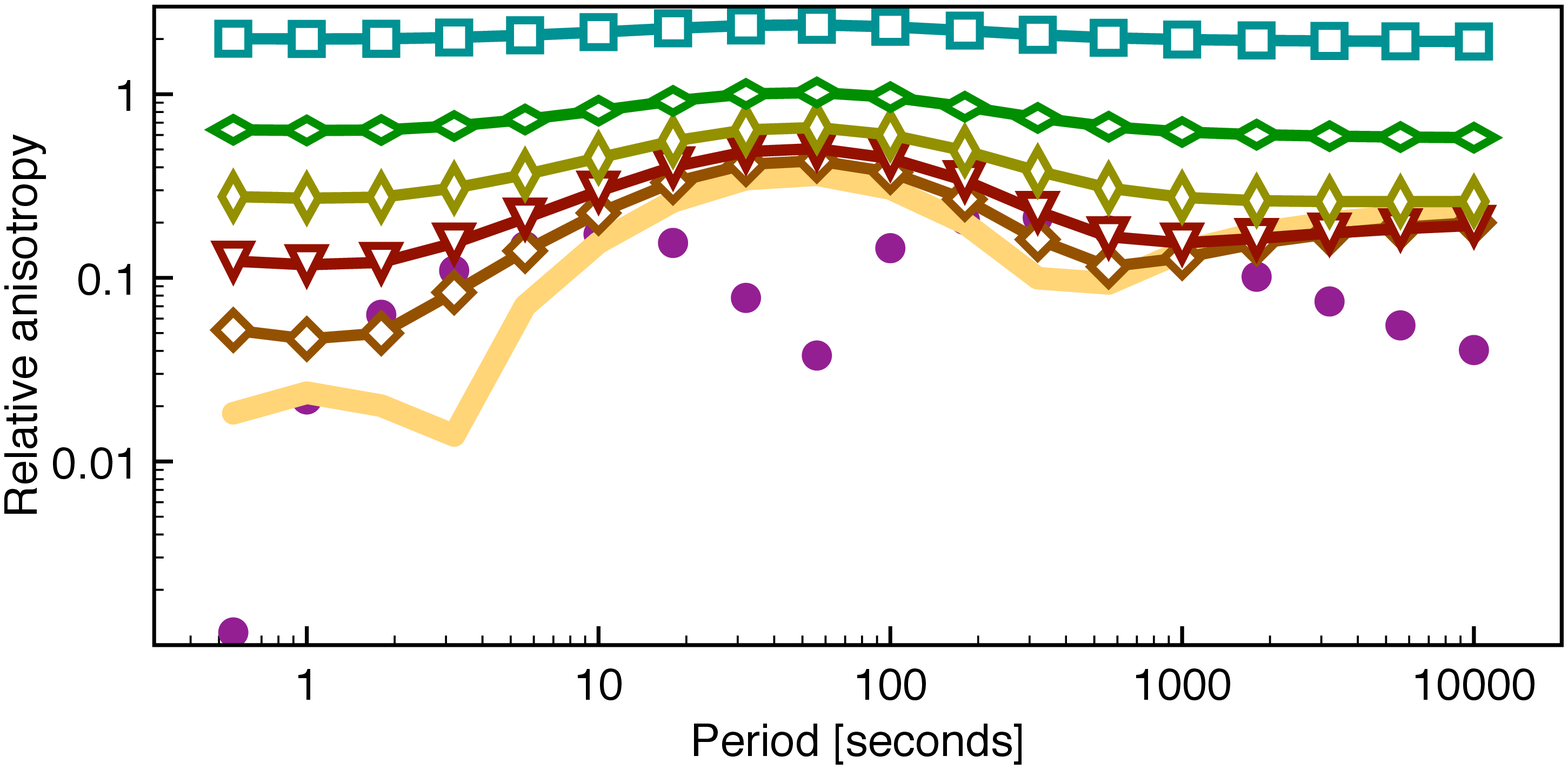}\\
		b) Mean of relative amplitude anisotropy as a function of twist, shear and anisotropic distortion.\\
		\includegraphics[width=.325\textwidth]{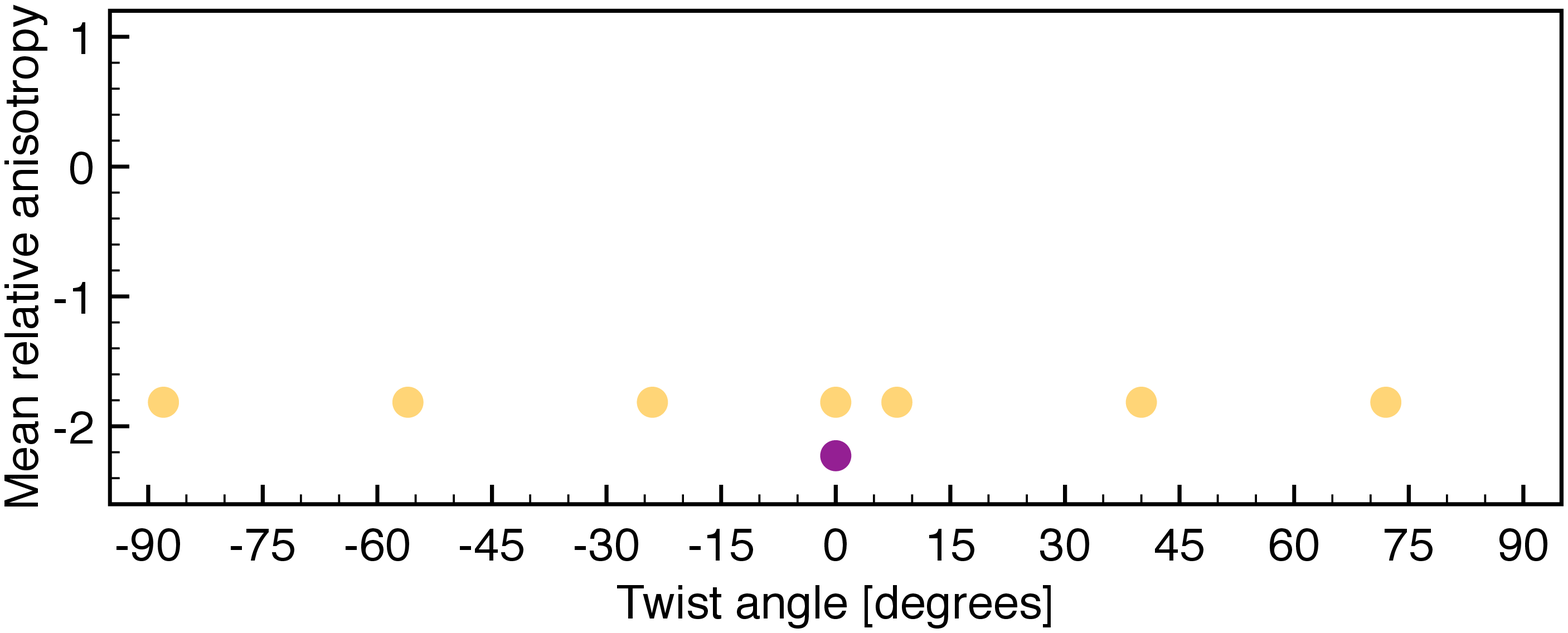}
		\includegraphics[width=.325\textwidth]{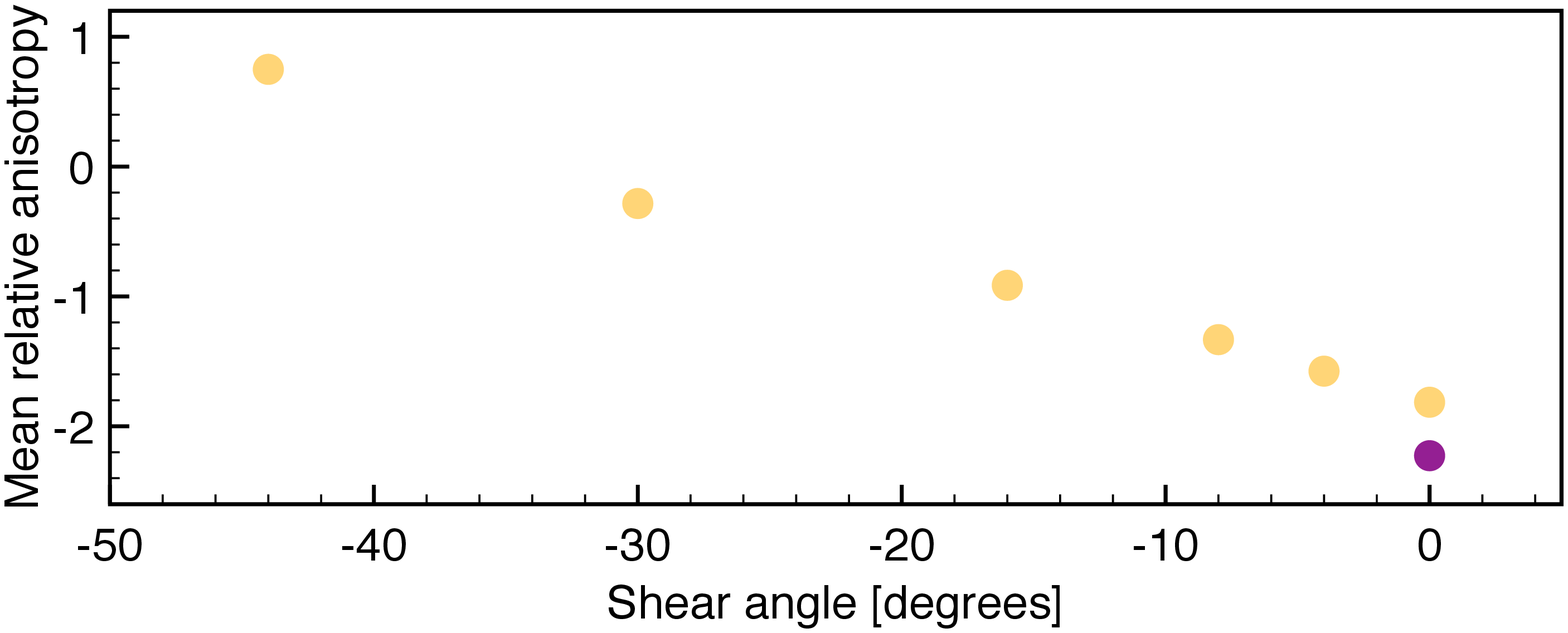}
		\includegraphics[width=.325\textwidth]{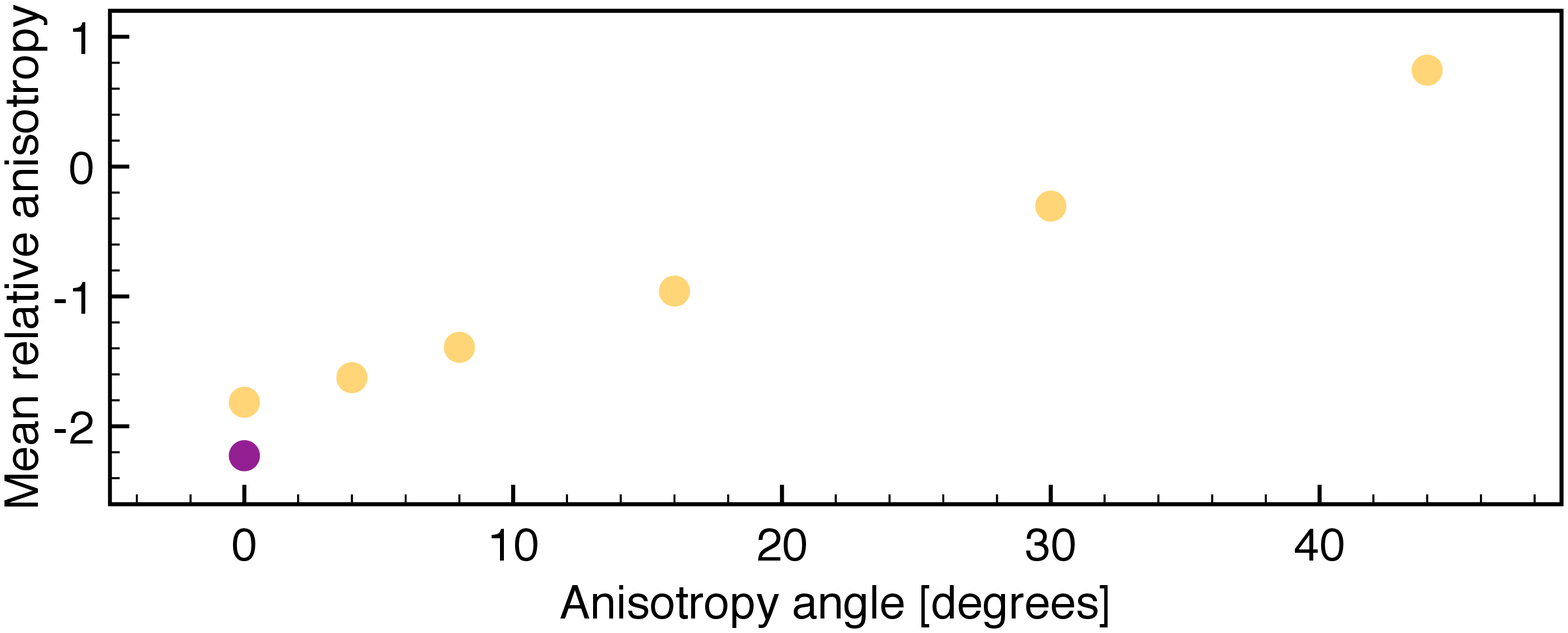}
	\caption{This figure illustrates how the Amplitude Tensor parameters are affected by distortion. The synthetic data is taken from the DSM1 \citep[site A09;][]{Miensopust:2013}. Each of the distortion parameters has been varied with the other two being fixed at no distortion; in the left column only the twist angle varies; in the middle column, the shear angle varies and in the right column varies the anisotropy angle. Shear and anisotropy angles are only plotted for negative and positive angles, respectively. Note, that twist distortion only affects the rotational parameters, skew and strike angles, but those are affected across the entire period range in contrast to shear and anisotropic distortion. Further note that shear and anisotropic distortion affect the singular values in a very similar fashion, which results in an increasing deviation of the relative anisotropy of the amplitude versus the relative anisotropy of the phase for increasing distortion. Keeping in mind the $90^{\circ}$ strike ambiguity, another important observation is that high shear (anisotropy) distortion approaches a strike angle of $45^{\circ}/135^{\circ}$ ($0^{\circ}/90^{\circ}$), because the electric field is strongly deflected (amplified) in one direction.}
	\label{fig:dA9}
\end{figure}
Let us discuss the effect of galvanic electric distortion on the Amplitude Tensor (AT) parameters on a brief example, the synthetic, noise-free data of site \emph{A09} from the \emph{DSM1} \citep{Miensopust:2013}. Galvanic distortion is applied systematically to
the data by varying independently each of the distortion parameters, $\phi_t$,
$\phi_s$ and $\phi_a$. All distorted data are subsequently analysed and plotted in
Figure \ref{fig:dA9} for comparison. The parameter study intents to show how
each AT parameter reacts on each distortion parameter.

The twist angle, $\phi_t$, is set to $-88^{\circ}$, $-56^{\circ}$, $-24^{\circ}$, $8^{\circ}$, $40^{\circ}$ and $72^{\circ}$, and clearly acts as an additional rotation \citep[as discussed by][]{Groom:1989} on the amplitude skew and strike angles. This is not a very surprising result, because the distorting twist rotation acts from the left on the decomposition of the AT (compare with \eqref{eq:TensorParameter} for a general tensor) and thus, shifts the left strike angle by the amount of the twist angle. Then, this shift of the strike angle must be counteracted by a shift of the skew angle to keep left and right strike angles
coherent. This explains why the skew angle is affected equally to the strike angle and why the singular
values and thus the relative anisotropy is unaffected by a sole twist distortion. In conclusion, twist distortion deviates skew and strike angles at all periods.

The shear angle, $\phi_s$, is set to $-0.5^{\circ}$, $-1^{\circ}$, $-2^{\circ}$, $-4^{\circ}$, $-8^{\circ}$, $-16^{\circ}$, $-30^{\circ}$ and $-44^{\circ}$, to illustrate characteristic behaviour for low and high shear. The positive angles are omitted but behave similar as they mirror the negative angles on the undistorted skew and strike angles, due to the inherent symmetry of the skew matrix, $\mathbf{S}$. Shear distortion disperses the skew angle
depending on how large the skew is, however, we can also observe two intersection
points where all curves coincide of which the second one is shared with the strike angle and can be explained physically. A similar situation is displayed for the strike
angle, where we find one intersection point and the dispersion
related to the inverse amplitude of the skew angle. For very high shear distortion, the strike angle assumes a constant value of $45^{\circ}$, reflecting the fact that both horizontal electric field components cannot be distinguished and point to $45^{\circ}$ (instead of $0^{\circ}$ and $90^{\circ}$) in local coordinates. The crossing of the real amplitude strike angle with $45^{\circ}$ represents the aforementioned intersection point, which exists because the strike angle converges to $45^{\circ}$ for increasing shear distortion.
Additionally, it appears that there is a large deviation for small shear angles when the discrepancy between singular values is small, because then the effect of the shear (approaching a strike of $45^{\circ}$) is more dominant. 
The singular values of the AT are affected by shear and we observe that the
difference between the singular values tends to always increase for increasing shear
angles, regardless of the sign of the shear angle (not shown here). This is reflected by an
apparent increase of mean and average relative anisotropy for an increasing
absolute shear angle. Summing up these observations, the shear distortion
amplifies anisotropy and skew, and heavily pulls the strike angle towards $45^{\circ}$.

The anisotropy angle, $\phi_a$, ranges over $0.5^{\circ}$, $1^{\circ}$, $2^{\circ}$, $4^{\circ}$, $8^{\circ}$, $16^{\circ}$, $30^{\circ}$ and $44^{\circ}$. Note that the matrix $\mathbf{A}$ is symmetric and therefore we only show positive angles. The AT skew angle is increasingly dispersed
with increasing period and increasing anisotropic distortion, leaving the low
periods  barely affected. The distorted AT strike angle
tends towards $0^{\circ}$ ($90^{\circ}$) increasing positive (negative) anisotropy, due to an increasing amplification of one electric field component over the other to the end that the dominant component constitutes the strike angle. The anisotropic distortion acts as a multiplicative factor on the singular values and thus shifts the curves up and down. The computation of the relative anisotropy reveals that the mean and average relative anisotropy are more indicative than the actual
singular values. In fact, this measure suggests (as for the shear) that the
relative anisotropy difference between AT and Phase Tensor (PT) should be
small if there is no distortion present in the data. Conclusively, the anisotropic distortion disperses the skew angle (more with increasing period), distorts the strike angle (towards the dominating component) and modifies
anisotropy (resulting in amplification of the difference of relative anisotropy
between phase and amplitude).

As a general conclusion to Figure \ref{fig:dA9}, we find that the AT parameters not
only behave similar to the PT parameters over a range of periods, but
also offer a diagnostic which allows to tell apart galvanic distortion of the electric field from the inductive-galvanic response of the measured impedance. Especially the twist angle displays very clear distortion signatures on amplitude skew and strike angles facilitating the comparison of similarity to the phase skew and strike angles. The parameters skew angle, strike angle and relative anisotropy show strong similarity between PT and AT, which diverges towards long periods. We explain this divergence by the accumulation of galvanic charges of increasingly large subsurface features with increasing period, which is part of the modelled EM response. 
Its removal (the entire short period galvanic response) from the AT results in a purely inductive AT at short periods with a natural, continuous build-up of the galvanic response with increasing period, that only originates from structures of the size comparable to the smallest inductive scale captured by the data and not smaller galvanic scales as, in contrast, contained in galvanic distortion.

\section{Considerations for the Weights in the Objective Function}
\label{ConsiderationsObjectiveFunction}
We realised the objective function as a weighted sum of four sub-objective functions as described in section \ref{sec:ComparisonFunction} and, in particular in \eqref{eq:FitnessFunction}, but to choose the appropriate weights of each objective in such a multi-objective function optimisation is not a trivial problem. The most general solution to this multi-objective optimisation results in a Pareto front, a set of equally valid solutions, of which none can be determined as best solution without further constraints. To choose one solution from the Pareto front, we propose to weight the individual sub-objectives by the inverse of the frequency-weighted variances, ${\hat{\sigma}}_{\psi_\Phi/\theta_\Phi/\alpha_\Phi}^2$, of the respective Phase Tensor parameters skew angle $\psi_\Phi$, strike angle $\theta_\Phi$ and anisotropy $\alpha_\Phi$:
 \begin{equation}
{\hat{\sigma}}_{\psi_\Phi/\theta_\Phi/\alpha_\Phi}^{-2} = \sum_i \frac{f_i^2}{\mathbf{ff}^*} \sigma_{\psi_\Phi/\theta_\Phi/\alpha_\Phi,i}^{-2}=\sum_i w_{\psi_\Phi/\theta_\Phi/\alpha_\Phi,i}.
\end{equation} 
The advantage of such weights is that they define a weighted single-objective optimisation \eqref{eq:FitnessFunction}, instead of much slower multi-objective functions without a definite solution, and that these weights adapt dynamically to the impedance data confidence by prioritising the sub-objective function that contains the better resolved parameters. Naturally, but at a high price in computation time, the exact same result can be obtained from the Pareto front of a multi-objective optimisation that treats each sub-objective function as an independent objective function, where the single-objective result would correspond to the Pareto result with the smallest weighted sum according to the data specific weights. 

Note that since $\det(\mathbf{W}_{\psi_\Phi/\theta_\Phi/\alpha_\Phi})=\hat{\sigma}_{\psi_\Phi/\theta_\Phi/\alpha_\Phi}^{-2}$, the formulations of the weight matrices $\mathbf{W}_{\psi_\Phi/\theta_\Phi/\alpha_\Phi}=\mathrm{diag}\left(w_{\psi_\Phi/\theta_\Phi/\alpha_\Phi,i}\right)$ in \eqref{eq:FitnessFunction} already include this sub-objective weighting proposition.



%

\end{document}